\documentclass[aps, prd, superscriptaddress, preprintnumbers, 
nofootinbib]{revtex4}
\usepackage{graphicx}
\usepackage{amsmath}
\def\fsl#1{\setbox0=\hbox{$#1$}                 
   \dimen0=\wd0                                 
   \setbox1=\hbox{/} \dimen1=\wd1               
   \ifdim\dimen0>\dimen1                        
      \rlap{\hbox to \dimen0{\hfil/\hfil}}      
      #1                                        
   \else                                        
      \rlap{\hbox to \dimen1{\hfil$#1$\hfil}}   
      /                                         
   \fi}                                         %

\newcommand{\VEV}[1]{\langle #1 \rangle}

\newcommand{\lessim}{\mathop{<}\limits_{\displaystyle{\sim}}}

\newcommand{\NDA}{\Omega_{\rm NDA}}

\newcommand{\TeV}{\text{TeV}}
\newcommand{\GeV}{\text{GeV}}
\newcommand{\NKKg}{N_{\rm KK}^g}
\newcommand{\NKKb}{N_{\rm KK}^{gs}}
\newcommand{\NKKf}{N_{\rm KK}^f}
\newcommand{\NKKs}{N_{\rm KK}^h}
\newcommand{\NKKi}{N_{\rm KK}^i}

\newcommand{\Li}{\Lambda^{-1}}
\begin{document}
\title{Top Mode Standard Model in Six Dimensions}
\thanks{
A preliminary version~\cite{DSB} was given at the 2004 International Workshop on Dynamical 
Symmetry Breaking (DSB 04), Dec. 21-22 , 2004, Nagoya University.
}

\date{\today}

\preprint{DPNU-05-13}

\author{Hidenori Fukano}
\email[E-mail: ]{fukano@eken.phys.nagoya-u.ac.jp}
\affiliation{Department of Physics, Nagoya University, Nagoya 464-8602, Japan}
\author{Koichi Yamawaki}
\email[E-mail: ]{yamawaki@eken.phys.nagoya-u.ac.jp}
\affiliation{Department of Physics, Nagoya University, Nagoya 464-8602, Japan}

\begin{abstract}
We construct a top-mode standard model where the third 
generation fermions and the $SU(2)_L \times U(1)_Y$
gauge bosons are put on a 6-dimensional brane (5-brane)
with the extra dimensions compactified on the TeV scale 
($R_5^{-1}=R_6^{-1} \equiv R^{-1}= 1-10 \,{\rm  TeV}$), while only
the gluons live in a compactified 8-dimensional bulk
($R_7^{-1}=R_8^{-1} \equiv \Lambda \gg R^{-1}$).
On the 5-brane,  Kaluza-Klein (KK) modes
of the bulk gluons give rise to induced four-fermion interactions 
which, combined with the  gauge interactions, 
are shown to be strong enough to trigger the top quark condensate,
based on the dynamics of  6-dimensional gauged Nambu-Jona-Lasinio (NJL) model. 
Moreover, we
can use a freedom of the brane positions to tune the four-fermion coupling close to the critical line of 6-dimensional gauged NJL model,
so that the gap equation can ensure the top condensate on the weak scale 
while keeping other fermions massless. There actually exists a scale 
(``tMAC scale''), 
$\Lambda_{\rm tM} = (7.8-11.0) R^{-1}$, where the running gauge couplings
combined with the induced four-fermion interactions trigger 
only the top condensate while no
bottom and tau condensates.  Furthermore,
presence of such explicit four-fermion 
interactions enables us to formulate straightforwardly 
the compositeness conditions at $\Lambda=\Lambda_{\rm tM}$, 
which, through the renormalization-group analysis, yields a prediction of
masses of the top quark and the Higgs boson, $m_t = 177 - 187 \,\GeV$ 
and $m_H = 183 - 207 \,\GeV$.
\end{abstract}

\maketitle

\section{Introduction}
The origin of mass is one of the most urgent problems in the modern particle 
physics.
The Standard Model (SM) has a mysterious part, the electroweak symmetry breaking (EWSB), 
to give mass to the elementary particles.
The EWSB via the elementary Higgs boson in the SM has many problems, 
fine-tuning problem, etc.
Particularly, the SM does not tell us why only the top quark 
has a mass of order
of the EWSB scale.

A simple solution was actually proposed much earlier than the
discovery of the top quark with the mass being this large, namely
the idea of top quark condensation
which was
proposed by Miransky, Tanabashi and Yamawaki (MTY)~\cite{MTY89}, 
based on the phase structure
of the gauged Nambu-Jona-Lasinio (NJL) model~\cite{KMY89,ASTW}, and independently by Nambu~\cite{Nambu89} 
in a different context (see also \cite{Marciano89}). 
In order to trigger the top quark condensate
$\langle \bar{t}t \rangle$, 
MTY introduced explicit four-fermion interactions: 
\begin{equation}
  {\cal L}_{4f}
   =G_t (\bar\psi_L^{i} t_R)({\bar t}_R {\psi_L}_i) + G_b (\bar\psi_L^{i} b_R)({\bar b}_R {\psi_L}_i ) 
    + G_{tb}\left(\epsilon^{ik}\epsilon_{jl}\bar\psi_L^i {\psi_R}_j
     \bar\psi_L^k {\psi_R}_l\right)
   + h.c.,
\label{eq:tb-4fermi}
\end{equation}
with $\psi_L=(t,b)_L$, and similarly for other generations.
The dimensionless four-fermion couplings $g_t$ is defined as 
$G_t\equiv  g_t (4\pi^2)/[N_c \Lambda^2]$ and similarly for $g_b$  and $g_{tb}$.
~\footnote{
In terms of the notation $g^{(1)},g^{(2)},g^{(3)}$ in Ref.~\cite{MTY89}, 
these couplings read
$g_t=g^{(1)} + g^{(3)}, g_b=g^{(1)}-g^{(3)}, g_{tb}=g^{(2)}$. 
$g_{tb}$will be disregarded for 
the moment. We shall come back to it in Sec.\ref{summary}} 
The situation $m_t \gg m_b, m_c$ is realized as the critical phenomenon,
$m_t \ne 0$ while $m_b=m_c=\cdots=0$ as the first approximation. This 
takes place when
\begin{equation}
 g_t> g^{\rm crit} >g_b,\, g_c, \cdots
\label{eq:tbsplit}
\end{equation}
where $g=g^{\rm crit}$ is the value on the critical line of second order phase transition
of the gauged NJL 
model:~\cite{KMY89,ASTW}
\begin{equation}
 g=g^{\rm crit}=\frac{1}{4}\left(1+\sqrt{1-\frac{\alpha}{\alpha^{\rm crit}}}\right)^2 
\label{criticalline4}
\end{equation}
with $\alpha=g^2/(4\pi)$ ($g$: gauge coupling const.) and $\alpha^{\rm crit}=\pi/3$. 
The gap equation (improved ladder Schwinger-Dyson (SD) equation)
dictates that the top mass can be 
much smaller than the cutoff $m_t \ll \Lambda$ 
by tuning the four-fermion coupling 
arbitrarily close to the critical line $0 <g_t-g^{\rm crit} \ll 1$. 
The model predicted a top mass of weak scale order
by the Pagels-Stokar formula~\cite{Pagels:hd} 
evaluated through the
solution of the gap equation and also
predicted a
scalar bound state $\bar{t}t$ which plays the role of the Higgs boson in the SM.
Thus the model was called  ``top mode standard model"(TMSM).
The TMSM was further formulated in an elegant fashion by Bardeen, Hill and 
Lindner (BHL)~\cite{BHL:1989ds} through the renormalization-group equations (RGE's) of the SM combined 
with the compositeness condition. (For reviews of TMSM
see \cite{Yamawaki96,Miransky:vk,Hill:2002ap}.)

However, the original TMSM has a few problems: The model needs ad hoc 
four-fermion  interactions whose origin is not known. Furthermore,
even if we assumed the cutoff, $\Lambda$, is the Planck scale, the predicted mass of the top quark is $m_t = 220-250 \,\GeV$, 
somewhat larger than the experimental value~\cite{MTY89,BHL:1989ds}.
If we avoided the fine-tuning  
by assuming $\Lambda$ is a few $\TeV$, then we would
get a disastrous prediction
$m_t \sim 600 \,\GeV$. 

As to the origin of the four-fermion interactions, an immediate possibility
is the massive vector boson exchange~\cite{Tanabashi:1989sz} 
whose explicit model was given by the
topcolor~\cite{Hill:1991at}
where the massive vector bosons (colorons) are the gauge 
bosons of the spontaneously broken extra gauge symmetry (topcolor) $SU(3)_1 \times SU(3)_2 \to SU(3)_c$. 
Further on this line, a solution to 
the top mass problem was given by the
top quark seesaw model (TSS)~\cite{TSS:1997nm,He:2001fz} 
where a new vector-like massive $SU(2)_L$-singlet
$\chi$-fermion (seesaw partner of the top quark) mixed with the $t_R$ 
pushes down the top mass.  
Note that the $S$ and $T$ parameter constraints from
the precision electroweak experiments are quite insensitive to 
introduction of massive vector-like fermions which contribute to the 
electroweak symmetry breaking~\cite{Maekawa:1994yd}.  
Then there arise new questions:
How does the topcolor symmetry breaking pattern occur? 
Where does the 
$\chi$-fermion  come
from?  

As an attractive answer to the above questions, the SM (without Higgs fields)
was embedded into higher dimensions 
with compactified extra dimensions~\cite{Dobrescu:1998dg,CDH:1999bg}:
The gluon Kaluza-Klein (KK) modes play the role of the topcolor yielding the top-mode four-fermion interactions,\footnote{
For the topcolor scenario with extra dimensions, see Ref.\cite{Hashimoto:2004xz}.
} 
while the KK modes of the $t_R$ 
(vector-like massive fermions) playing the role of the $\chi$-fermion.
Then the TMSM with compactified extra dimensions is essentially equivalent to 
TSS, although the diagonal mass $\bar{t} t$ does exist in contrast to 
the seesaw mechanism.

A more straightforward version of the TMSM with extra dimensions
was proposed by Arkani-Hamed, Cheng, Dobrescu and Hall (ACDH)~\cite{ACDH:2000hv}:
All the third generation fermions and the SM gauge bosons are put in the
$D (=6,8,10,\cdots)$-dimensional bulk on the equal footing, while other fermions 
are fixed on the 3-brane. Note that the four-fermion couplings are totally 
replaced by the bulk SM gauge dynamics and hence are no longer freely adjusted. 
Although it was expected in Ref.~\cite{ACDH:2000hv} 
that the pure gauge dynamics 
of the bulk SM (without four-fermion interactions)  may give rise to the top 
quark condensate, it was found~\cite{HTY:2000uk,Gusynin:2002cu},  
that the bulk QCD in six 
dimensions actually is not strong enough as to trigger the top quark condensate
within 
the analysis based on the truncated KK effective theory~\cite{DDG} for the
running of gauge couplings and the 
 (improved) ladder Schwinger-Dyson (SD) equation.\footnote{
See Refs. \cite{Rius:2001dd} for other scenarios based on the  
Randall-Sundrum type extra dimensions, which can yield enhanced condensates. 
 }
One might hope that including the bulk $U(1)_Y$ gauge interaction would enhance 
the attractive forces strong enough to trigger the top quark condensate. However,
it was  shown~\cite{HTY2003} 
by the full analysis of the bulk SM including the bulk
$SU(2)_L\times U(1)_Y$ gauge interactions that the $D=6$ 
version of the ACDH scenario does not realize 
the ``topped MAC'' (tMAC) where the
top quark condensate is the most attractive channel (MAC),  
since then the strong $U(1)_Y$ interaction favors the tau condensate 
rather than the top quark condensate. Giving up the possibility of $D=6$ case,
the tMAC analysis in Ref.~\cite{HTY2003} showed that $D=8$ 
version yields a viable model, predicting the mass of the top quark and the
Higgs boson, $m_t = 172-175 \,{\rm GeV}, m_H = 176-188 \,{\rm GeV}$.

Is there no chance for the $D=6$ TMSM to survive, then?
Quite recently, phase structure of the $D (=6,8,10,\cdots)$-dimensional 
gauged NJL model was revealed~\cite{Gusynin:2004jp}, which is
similar to the four-dimensional one~\cite{KMY89}, Eq.(\ref{criticalline4}). 
The result
suggests that the top quark condensates even for $D=6$ TMSM 
can be formed thanks to the additional
four-fermion interactions as in the original 
TMSM in four dimensions. Although it might be a kind of drawback
to introduce arbitrary ad hoc four-fermion interactions, here we instead follow
the method of  Ref.~\cite{Dobrescu:1998dg,CDH:1999bg} to {\it generate such  
four-fermion interactions} on the 5-brane {\it out of the gauge dynamics} in 
higher dimensional bulk, namely
the KK modes of the bulk gluons in 8 dimensions with 
compactified extra dimensions.  

In this paper, based on the phase structure of the 6-dimensional gauged NJL 
model~\cite{Gusynin:2004jp}, we shall construct a TMSM on the 5-brane  
with the third generation 
fermions and the $SU(2)_L\times U(1)_Y$ gauge bosons 
living on 
the  5-brane with the extra dimensions compactified on $T^2/Z_2$ with  TeV scale 
($R_5^{-1}=R_6^{-1} \equiv R^{-1}= 1-10 \,{\rm  TeV}$), while
only the gluons live in the  8-dimensional bulk compactified on  $T^2/Z_2$,
($R_7^{-1}=R_8^{-1}\equiv \Lambda \gg R^{-1}$), where $R_{D}$ are the radii of 
D-dimensional compactification. 
Fermions of the first and the second generations are fixed on the 3-brane.
We shall show that the induced four-fermion coupling  indeed exceeds the 
value on the critical line of the gauged NJL model for the top condensate to take place. 
Moreover, we have a freedom of choosing the brane position by exploiting 
the compactification based on  $T^2/Z_2$ rather than $T^2$, which
is crucial for two 
reasons: The top quark mass must be kept to be of weak scale via the SD gap equation
by tuning the four-fermion coupling
close to the critical line (given the values of the SM gauge couplings on the brane).  
In order to realize the top condensate but not the bottom condensate,
we further need to tune the four-fermion coupling close to the critical line
in such a way that the
$U(1)_Y$ gauge interaction discriminates the top in the broken phase 
(above the critical line) and the bottom in the 
symmetric phase (below the critical line).
Important point is that what is uniquely dictated by the bulk QCD coupling is the
upper bound
of the induced four-fermion coupling on the 5-brane, whereas the actual value of it  
can be tuned arbitrarily smaller than its upper bound thanks to a 
freedom of the brane positions of our compactification $T^2/Z_2$.  Thus, {\it as far as the
upper bound exceeds the value on the critical line,
 we can tune the brane positions
so that the SD gap equation 
can ensure the top mass on the weak scale much smaller than} 
$\Lambda$, $m_t \ll \Lambda$,~\footnote{
In the MTY formulation~\cite{MTY89}, the gap equation yields a relation 
between $m_t$, $\Lambda$ and the four-fermion 
coupling, while the PS formula does a relation between $m_t$ and
$\Lambda$ once we fix the weak scale $F_\pi=246 \,{\rm GeV}$. The PS formula 
yields a realistic top mass $m_t \ll \Lambda$, which can be compatible to
the gap equation only when the four-fermion
coupling have a freedom to be tuned close to the critical point.
A similar consistency requirement exists 
also in the equivalent formulation of BHL~\cite{BHL:1989ds} 
where the RGE's combined with the compositeness condition play the roles of the
gap equation and the PS formula (see, e.g. \cite{Yamawaki96}).
}
while keeping other fermions massless (as a zeroth approximation).
We actually find a tMAC scale $\Lambda_{\rm tM}$ where
  the running gauge couplings,
combined with the induced four-fermion interactions, trigger 
only the top condensate while no
bottom and tau condensates. The tMAC scale in this model reads 
$\Lambda_{\rm tM} = (7.8-11.0) R^{-1}$. Here we use the value of the critical 
binding strength in the {\it nonlocal gauge} in the SD equation~\cite{Gusynin:2002cu,Gusynin:2004jp}, which is larger than 
the (``conservative'') one used in the previous analysis~\cite{HTY2003} based 
on the 
the Landau gauge~\cite{HTY:2000uk}, and hence our conclusion on the existence of
the tMAC scale  is quite {\it independent of the
ambiguity in the SD equation analysis}. (In Ref.~\cite{HTY2003}
there would have been no tMAC scale even for $D=8$, 
if the value of the nonlocal gauge 
were used.)
 Moreover, in contrast to the previous models~\cite{ACDH:2000hv,HTY2003}, 
presence of such explicit four-fermion 
interactions enables us to formulate straightforwardly 
the compositeness conditions at $\Lambda= \Lambda_{\rm tM} $
which, through the RGE analysis {\' a} la BHL~\cite{BHL:1989ds}, yields a prediction of
masses of the top quark $m_t$ and the Higgs boson $m_H$, $m_t = 177 - 187 \, \GeV$ and $m_H = 183 - 207 \,\GeV$.

The paper is organized as follows. In Sec.\ref{model} we recapitulate 
the binding strength of the SM 
gauge couplings on the 5-brane.
In Sec.\ref{4fermi in 6d}, we derive four-fermion interactions on the 5-brane
induced out of 8-dimensional bulk gluons and estimate the strength of them. 
In Sec.\ref{gnjl in 6d}, we show that the induced  four-fermion couplings and the SM gauge couplings for the top quark 
on the 5-brane are
larger than the critical line value of the gauged NJL model 
in six dimensions in such a way that 
only the
top condensate takes place, 
while other fermions do not condense. Moreover
the freedom of the brane positions can be used
to tune the four-fermion coupling arbitrary close to the critical line
so that the gap equation keeps the top mass on the weak scale order. 
In Sec.\ref{predict mass}, based on the BHL procedure 
of the RGE's and the compositeness condition, we predict the 
masses of top quark and Higgs boson for the 6-dimensional TMSM.
Sec.\ref{summary} is devoted to summary and discussions.

\section{Binding strength of the Standard Model gauge couplings on the 5-brane}
\label{model}
Here we  depict the result of the one-loop running of the bulk gauge couplings of the  SM  
in the KK effective theory~\cite{DDG} used for
the tMAC analysis in Ref.~\cite{HTY2003} ($SU(2)_L$ gauge coupling is irrelevant to the binding strength for
the condensate).
First, one-loop RGEs of four-dimensional couplings QCD ($g_3$), $SU(2)_L$ 
($g_2$)
and $U(1)_Y$ 
($g_Y$)  below the compactification scale  $R^{-1}_D$ are given by
\begin{equation}
  (4\pi)^2 \mu \frac{d g_i}{d \mu} = b_i\,g_i^3, \quad (\mu < R^{-1}_D),
\label{rge_4d}
\end{equation}
where $b_3=-7, b_2=-19/6, b_Y=41/6$.
Above the compactification scale the RGEs of $D$-dimensional QCD, $SU(2)_L$ and $U(1)_Y$ couplings in the truncated KK effective theory~\cite{DDG} are given by  
\begin{equation}
  (4\pi)^2 \mu \frac{d g_i}{d \mu} = b_i\,g_i^3 
   + b_i^{\rm KK}(\mu)\,g_i^3, 
  \quad (\mu \geq R^{-1}_D), \label{rge-1}
\end{equation}
where $b_i^{\rm KK}$ for one generation and one (composite) Higgs boson are
\begin{eqnarray}
  b_3^{\rm KK}(\mu) &=& -11\, \NKKg (\mu)
 +\frac{\delta}{2}\, \NKKb (\mu) \nonumber \\ &&
 +\frac{8}{3}\, \NKKf (\mu), \label{bKK-qcd}
\end{eqnarray}
\begin{eqnarray}
  b_2^{\rm KK}(\mu) &=& -\frac{22}{3} \, \NKKg (\mu)
 +\frac{\delta}{3} \, \NKKb (\mu) \nonumber \\ &&
 +\frac{8}{3} \, \NKKf (\mu) \nonumber \\ &&
+ \frac{1}{6} \, \NKKs (\mu) \label{b2KK},
\end{eqnarray}
\begin{equation}
  b_Y^{\rm KK}(\mu) = 
 \frac{40}{9} \, \NKKf (\mu)
 + \frac{1}{6} \, \NKKs (\mu), \label{byKK}
\end{equation}
with $\delta \equiv D-4$ and $\NKKi(\mu) (i=g,gs,f,h)$ being the total numbers of KK modes below $\mu$ for gauge bosons, gauge scalars, 
four-component fermions and composite Higgs bosons, respectively.
We take the relation that $\NKKi(\mu)$ are 
\begin{equation}
 N_{\rm KK}^i (\mu) = \frac{1}{2^{\delta/2}}
                      \frac{\pi^{\delta/2}}{\Gamma(1+\delta/2)}(\mu R_D)^\delta. 
\label{NKK_app}
\end{equation}
The RGE's can be solved with the inputs from Ref.~\cite{PDG}: 
\begin{eqnarray}
  &&\alpha_3(M_Z)=0.1172 \, ,\label{qcd-mz} \\
  &&\alpha_2(M_Z)=0.033822 \, ,\label{su2-mz} \\
  &&\alpha_Y(M_Z)=0.010167, \label{u1-mz}
  \label{inputs}
\end{eqnarray}
where $\alpha_i (\equiv g_i^2/(4\pi))$ are the value at $\mu=M_Z(=91.1876 \;{\GeV})$.

We relate the four-dimensional gauge coupling, $g_i$, to the $D$-dimensional gauge coupling, $g_{D,i}$, 
at the compactification scale: $R_D^{-1}$
for $T^{\delta}/Z_2^k$ as
\begin{equation}
 g_{D,i}^2 = \frac{(2 \pi R_D)^{\delta}}{2^k}g_i^2,
\end{equation}
and define {\it dimensionless} $D$-dimensional coupling: $\hat{g}_{Di}(\mu)$ as 
\begin{equation}
\hat{g}_{Di}^2(\mu) \equiv g_{Di}^2(\mu)\mu^{D-4}.
\label{dimlessgauge}
\end{equation}
Hence we obtain
\begin{equation}
 \hat{g}_{D,i}^2(\mu) = \frac{(2 \pi R \mu)^{\delta}}{2^k}g_i^2.
\label{gc-1}
\end{equation}

By Eq.(\ref{rge-1}), Eq.(\ref{NKK_app}) and Eq.(\ref{gc-1}), we find RGEs
for the dimensionless $D$-dimensional couplings:
\begin{equation}
 \mu \frac{d}{d \mu} \hat g_{D,i} = \frac{\delta}{2}\,\hat g_{D,i}
 + \left(1+\frac{\delta}{2}\right) \NDA \, b'_i\, \hat g_{D,i}^3, 
 \quad \mbox{for } \mu \gg R_D^{-1}, \label{rge_2}
\end{equation}
where 
\begin{equation}
 \NDA \equiv \frac{1}{(4\pi)^{D/2}\Gamma(D/2)},
\end{equation}
and 
\begin{eqnarray}
  b'_3 &=& -11+\frac{\delta}{2}+\frac{4}{3}\cdot 2^{\delta/2} ,\\
  b'_Y &=& ~~\frac{20}{9}\cdot 2^{\delta/2} + \frac{1}{6} .
\end{eqnarray}
(As noted before, the $SU(2)_L$ coupling is irrelevant to the condensate.)

Eq.(\ref{rge_2}) implies that there exists
a nontrivial ultraviolet fixed point (UVFP) for $\hat{g}_{D,i}$: 
$\hat{g}_{D*,i}$~\cite{HTY:2000uk} (see also \cite{Agashe:2000nk,Kazakov:2002jd,Dienes:2002bg}):\footnote{
Two-loop contributions make the value of UVFP smaller
in the case at hand and hence even favor the existence of UVFP.~\cite{HTY:2000uk}
}
\begin{equation}
  \hat{g}_{D*,i}^2 \NDA = \frac{1}{-(1+2/\delta)\,b'_i}, \label{UVFP1}
\end{equation}
for $b'_i < 0$.
 For $D=6$ case with the compactification $D=6 \rightarrow  D=4$ as $T^2/Z_2 (\delta=2, k=\delta/2=1)$, 
the UVFP of the six dimensional QCD coupling ($D=6,i=3$) is
\begin{equation}
 \hat{g}_{6*,3}^2 \NDA = \frac{3}{44}.
\label{UVFP}
\end{equation}

\begin{figure}
\begin{tabular}{cc}
\begin{minipage}{0.5\hsize}
\begin{flushleft}
  \hspace*{0.1cm} {(a) \hspace*{5mm} \underline{$R_6^{-1}=10\,\TeV$}}
\end{flushleft}
  \vspace*{-0.9cm}
\includegraphics[width=\hsize,clip]{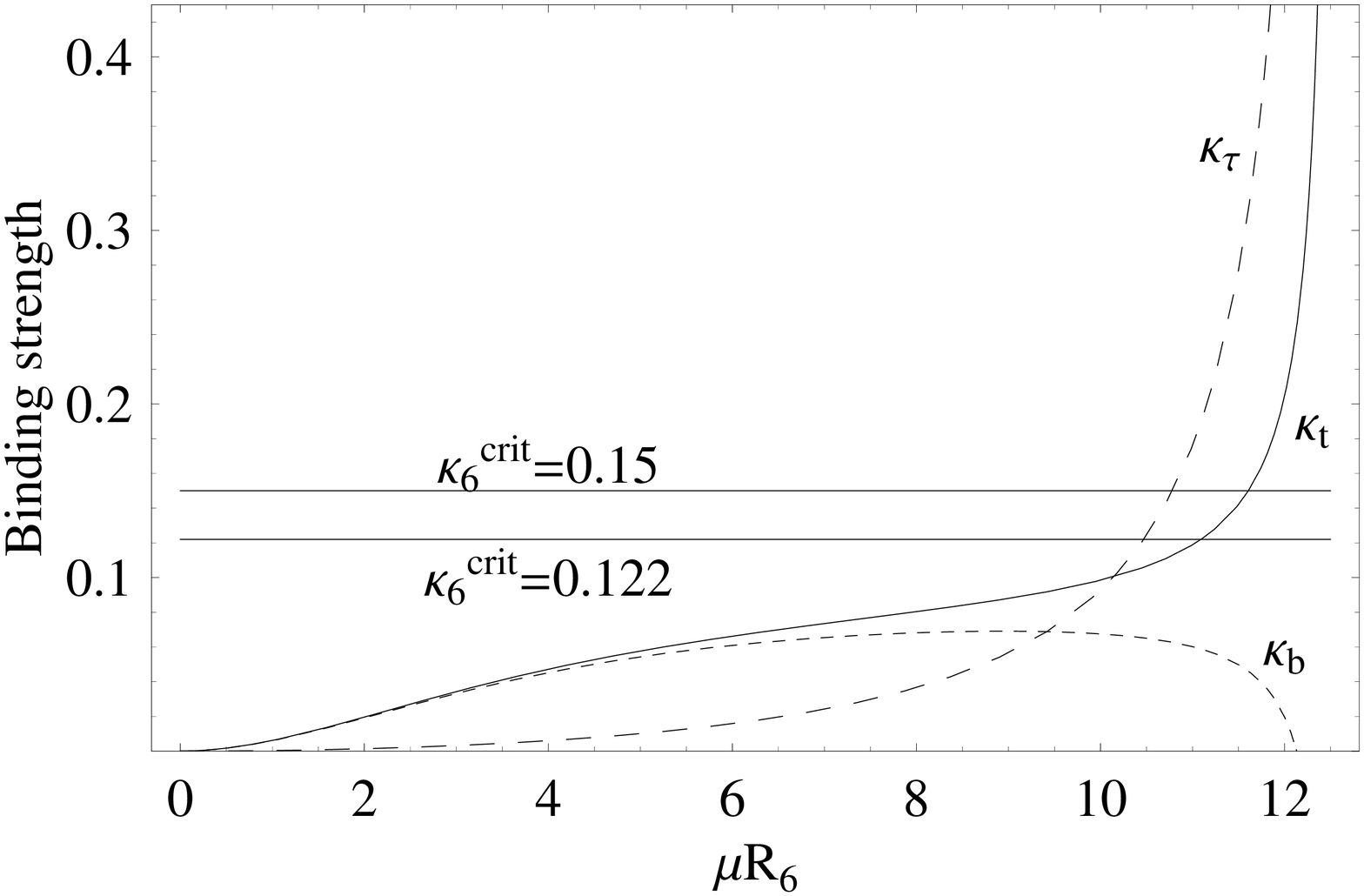}
\end{minipage}
\begin{minipage}{0.5\hsize}
\begin{flushleft}
  \hspace*{0.1cm} {(b) \hspace*{5mm} \underline{$R_6^{-1}=1\,\TeV$}}
\end{flushleft}
  \vspace*{-0.9cm}
\includegraphics[width=\hsize,clip]{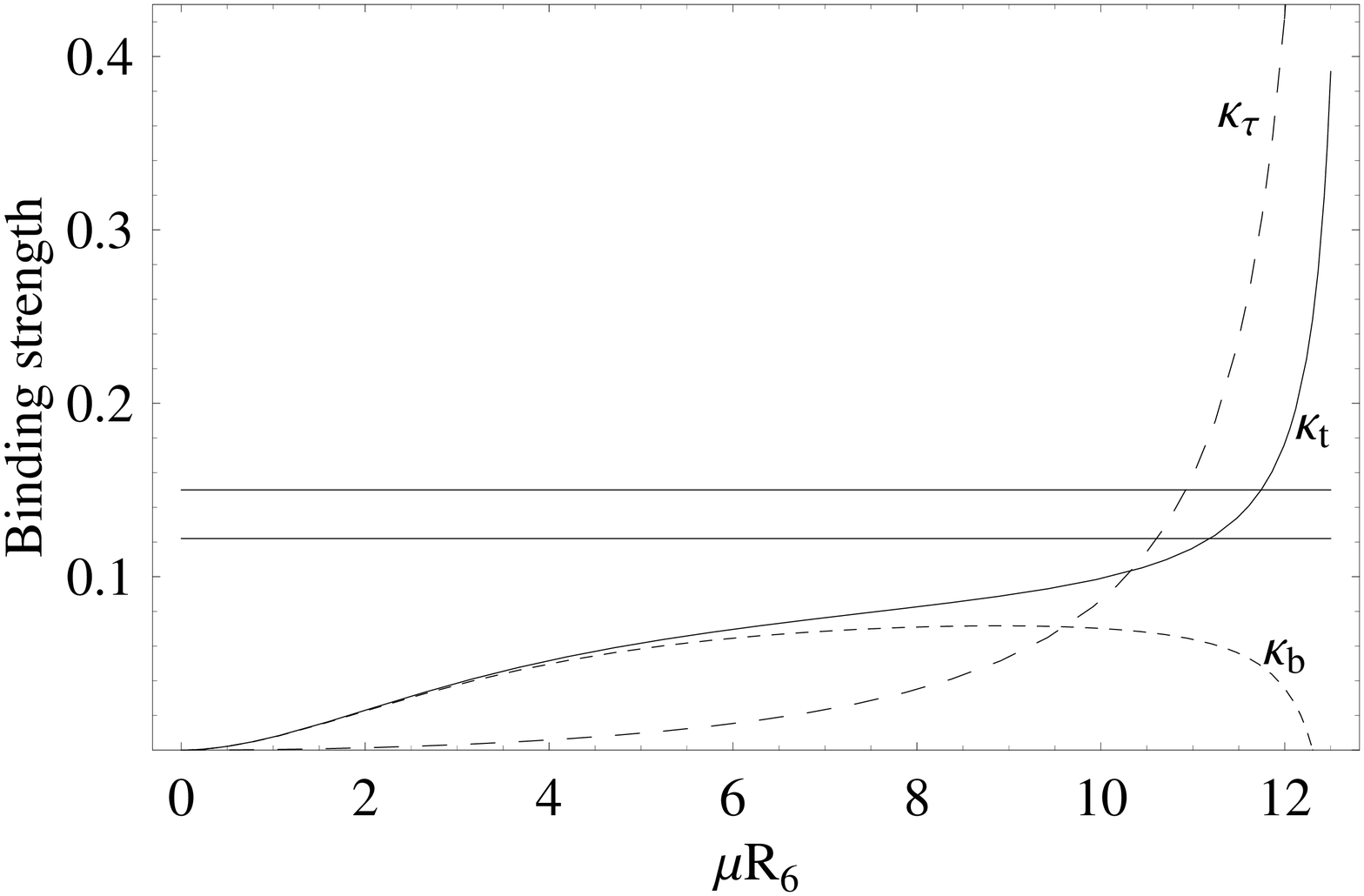}
\end{minipage}
\end{tabular}
\caption{Binding strengths $\kappa_{\alpha}$ ($\alpha=t,b,\tau$) for the top, bottom and $\tau$ on the 5-brane:
         A compactification scale is (a) $R_6^{-1}= 10\,\TeV$ and (b) $R_6^{-1}= 10
         \,\TeV$.
         Note that the upper horizontal line is $\kappa_6^{\rm crit}$ by using the nonlocal gauge fixing 
         method ($\kappa_6^{\rm crit}=0.15$) and
         the lower one is $\kappa_6^{\rm crit}$ by using the Landau gauge fixing method ($\kappa^{\rm crit}=0.122$)
         ~\cite{HTY2003, Gusynin:2002cu}.
         } 
\label{fig:tMAC0}
\end{figure}

Next, based on the one-gauge-boson-exchange approximation~\cite{Raby:1979my}, 
the binding strength of a scalar channel ($\bar{\psi}\chi$) is defined as
\begin{eqnarray}
 \kappa (\mu) &\equiv& 
 -\hat g_{D,3}^2 (\mu)\NDA \mbox{\boldmath $T$}_{\bar{\psi}} \cdot
                  \mbox{\boldmath $T$}_{\chi}  \nonumber \\ &&
 -\hat g_{D,2}^2 (\mu)\NDA \mbox{\boldmath $T'$}_{\bar{\psi}} \cdot
                  \mbox{\boldmath $T'$}_{\chi}  \nonumber \\ &&
 -\hat g_{D,Y}^2 (\mu)\NDA Y_{\bar{\psi}} Y_{\chi}, 
\end{eqnarray}
where {\boldmath $T$}, {\boldmath $T'$} are the generators of $SU(3)_c$, $SU(2)_L$ respectively, 
and {\boldmath $Y$} is the hypercharge.
{\boldmath $T$}, {\boldmath $T'$} fulfill 
\begin{equation}
 -\,\mbox{\boldmath $T$}_{\bar{\psi}} \cdot \mbox{\boldmath $T$}_{\chi}
 = \frac{1}{2}\left( C_2(\bar{\psi}) + C_2(\chi)
                    -C_2(\bar{\psi}\chi) \right),
\end{equation}
with $C_2(r)$ being the quadratic Casimir for the representation $r$ of the SM gauge group on the 5-brane.
Hence we calculate the binding strength $\kappa_\alpha (\mu)$ for each channel:
\begin{eqnarray}
  \kappa_{t} (\mu)&=& C_F \hat g_{D,3}^2 (\mu) \NDA 
                 + \frac{1}{9}\hat g_{D,Y}^2 (\mu) \NDA \label{k_t}, \\
  \kappa_{b} (\mu)&=& C_F \hat g_{D,3}^2 (\mu) \NDA 
                 - \frac{1}{18}\hat g_{D,Y}^2 (\mu) \NDA \label{k_b}, \\
  \kappa_{\tau} (\mu)&=& \phantom{C_F \hat g_{D,3}^2 (\mu) \NDA + \;\;\;}
                   \frac{1}{2} \hat g_{D,Y}^2 (\mu) \NDA \label{k_tau},
\end{eqnarray}
for the $D-$dimensional top, bottom and tau condensate, respectively and
$C_F=4/3$ is the quadratic Casimir of the fundamental representation of $SU(3)_c$. Note again that $SU(2)_L$ gauge interactions are opearative only on the left-handed
fermions and hence do not contribute to the biding strength for the scalar 
channel. In Fig.\ref{fig:tMAC0} 
the resultant running of the binding strengths in 
Ref.\cite{HTY2003} is depicted.

By using the improved ladder SD equation for the 
pure gauge dynamics on the 5-brane, 
Ref.~\cite{HTY:2000uk, Gusynin:2002cu} estimated that the critical binding strength $\kappa_D^{\rm crit} (D=6)$ is 
\begin{eqnarray}
\kappa_6^{\rm crit}
\left\{ 
\begin{array}{ll}
 \simeq & 0.122 \hspace*{2mm} (\text{the Landau gauge fixing}), \\
 = & 0.15 \hspace*{3.8mm} (\text{the nonlocal gauge fixing}). \\
\end{array} 
\right.
\label{criticalkappa}
\end{eqnarray}
Condensation takes place in the channel where
the $\kappa_\alpha(\mu)$ ($\alpha=t,b,\tau$) exceeds the critical value at certain $\mu$. 
When we increase the energy scale $\mu$, the dimensionless couplings and hence $\kappa_\alpha$ grow 
so that the $\kappa_\alpha$ in 
the MAC at certain point exceeds the 
$\kappa_6^{\rm crit}$. The Landau gauge estimate yields a value of $\kappa_6^{\rm crit}$
smaller than that of the nonlocal gauge and hence was used in Ref.
\cite{HTY2003}
as a conservative criterion for
the top condensate. 
Shown by Fig.~\ref{fig:tMAC0}, $\kappa_t$ in the top channel
does not exceed  the critical binding strength before the tau channel $\kappa_\tau$
does, even if we exploited
a conservative estimate of the Landau gauge fixing method.
Hence, it was concluded~\cite{HTY2003} for $D=6$  that within the pure gauge dynamics there is no energy scale region where the top quark condensate is the MAC (tMAC scale).

In what follows we shall consider a new situation where the induced four-fermion interactions arising from the bulk gluon interaction in addition to the gauge interactions on the 5-brane can give rise to existence of tMAC, {\it even if we exploit the nonlocal gauge} estimate of the $\kappa_6^{\rm crit}$. Actually, since the Landau gauge in the improved ladder SD
equation has a problem with the 
chiral Ward-Takahashi identity\cite{Kugo:1992pr}, 
we here use the nonlocal gauge value.  Thus our conclusion of the existence of the tMAC scale will be independent of the ambiguity of the SD equation 
analysis on the
$\kappa_6^{\rm crit}$.

\section{Induced four-fermion interactions on the 5-brane}
\label{4fermi in 6d}

Following Ref.~\cite{CDH:1999bg}, we first show that four-fermion interaction on the 5-brane are induced 
by gluonic KK-mode exchanges.

First let us consider a QCD Lagrangian on the 5-brane 
with the gluons in the seven-dimensional bulk for 
illustration:
\begin{equation}
\mathcal{L}_{7D} = \delta(x_7 - x_{70}) \bar{\psi}(X) i \Gamma^{M} D_{M} \psi(X). 
\label{eq:7dlag-1}
\end{equation}
where, 
\begin{equation*}
 D_{M} = \partial_{M} - i g_{7,3} G^a_{M}(X, x_7) T^a,
\end{equation*}
with $M = (\mu,5,6), (\mu=0,1,2,3)$ and $X=(x,y), x=x_{\mu} , y=x_{5,6} $ and $T^a \, (a=1,2,\cdots)$ are $SU(3)_c$ indexes. 
The above Lagrangian is gauge-invariant on the 5-brane $(M = \mu,5,6)$.

In order to compactify the seventh dimension, we impose 
boundary conditions ($S^1/Z_2$-compactification): 
\begin{equation}
 G_M(X,x_7)= G_M(X,x_7+2 \pi R_7) \label{eq:7bc1}
\end{equation}
\begin{equation}
 G_M(X,x_7)= G_M(X,-x_7) \label{eq:7bc2}
\end{equation}

Hence the gluons in the seven-dimensional bulk are decomposed as follows.
\begin{equation}
 G_M (X , x_7) = \frac{1}{\sqrt{2 \pi R_7}} \Biggl[ G_{M,0}(X) 
 + \sqrt{2} \sum_{n = 1}^{n_{\rm KK}} G^{[n]}_{M,c}(X) \cos\frac{n x_7}{R_7} \Biggr].
\label{eq:7d-gkk1}
\end{equation}
Hereafter, we will call the gluons KK-modes ``colorons".

The induced four-fermion interactions on the 5-brane between the fermions on the 5-brane and colorons may be approximated by the
one-coloron exchanges of the KK tower.
By taking account of the brane position $x_{70}$, which represents where the 5-brane exists in the seven-dimensional bulk in the compactification of 
$S^1/Z_2$,  
we have an effective Lagrangian (gauged NJL model)
in the 5-brane:
\begin{equation}
\mathcal{L}_{6D} =\bar{\psi}(X) i \Gamma^{M} D_{M,0} \psi(X) + \mathcal{L}_{\rm 4F},
\label{eq:6dlag-GNJL01}
\end{equation}
where
\begin{eqnarray}
D_{M,0} &=&  \partial_{M} - i g_{6,3} G^a_{M,0}(X) T^a, \quad g_{6,3}^2=2\frac{g^2_{7,3}}{2 \pi R_7}\Bigg|_{\mu=\Lambda}, \\
 \mathcal{L}_{\rm 4F} &=& - \frac{2}{2 \pi R_7}{\sum_{n = 1}^{n_{\rm KK}} g^2_{7,3} (M_n) \frac{\cos^2({n x_{70}}/R_7)}{2 M^2_n}} 
(\bar{\psi} \Gamma^M T^a \psi)(\bar{\psi}\Gamma_M T^a \psi), \label{eq:6dlag-GNJL02}
\end{eqnarray}
where  $M_n=n \Lambda$ ($\Lambda \equiv R_7^{-1}$) is the $n$-th KK mode mass and
$n_{\rm KK}$ is the largest number of ``colorons" we take into account
(we implicitly assume 
that a bulk gauge theory is a cutoff theory, since it is unrenormalizable
in the usual sense. Were it not for the ultraviolet cutoff of the bulk, 
$n_{\rm KK}$ would be infinite).

After Fierz transformations, scalar and pseudoscalar channels in (\ref{eq:6dlag-GNJL02}) read
\begin{equation}
\begin{split}
 \mathcal{L}_{\rm 4F}^\prime &= \left[
 \frac{3}{4} \cdot \frac{1}{2} \cdot \frac{2}{2 \pi R_7} {\sum_{n = 1}^{n_{\rm KK}} g^2_{7,3}(M_n) 
\frac{\cos^2({n x_{70}}/R_7)}{2 M^2_n}}
\right] \cdot
 \Bigl[(\bar{\psi} \psi)^2 + (\bar{\psi} i \Gamma_A \psi)^2 \Bigr] \\
 &= \frac{G_6^{(7)}}{2 N_c} \cdot
 \Bigl[(\bar{\psi} \psi)^2 + (\bar{\psi} i \Gamma_A \psi)^2 \Bigr] ,
\end{split}
\label{eq:6dlag-GNJL03}
\end{equation}
where  the factors 3/4  and 1/2 are from Fierz transformation for Lorentz indices  and for  the $SU(3)_c$ generator, respectively,
$G_6^{(7)}$ is the six-dimensional induced four-fermion coupling  
from the seven-dimensional bulk gluons, and
$\Gamma_A\equiv \Gamma^0 \Gamma^1 \Gamma^2 \Gamma^3 \Gamma^4  \Gamma^5 \Gamma^6$ is the 6-dimensional chirality matrix.

The summation in the coefficient of the four-fermion operator in (\ref{eq:6dlag-GNJL03}) may be written as
\begin{equation}
\begin{split}
 \sum_{n=1}^{n_{\rm KK}} g^2_{7,3}(M_n) \frac{\cos^2({n x_{70}}/R_7)}{2 M^2_n} &= 
  \sum_{n=1}^{n_{\rm KK}} g^2_{7,3}(n\Lambda) \frac{\cos^2(nx_{70}/R_7)}{2 (n\Lambda)^2} \\
 &= \frac{2\pi R_7}{2}\frac{g_{6,3}^2(\Lambda)}{2 \Lambda^2} \sum_{n = 1}^{n_{\rm KK}} \frac{\cos^2({n x_{70}}/R_7)}{n^5},
\end{split} \label{eq:6dlag-GNJL04}
\end{equation}
by considering the running effect of bulk gauge coupling:
\begin{equation}
 g_{7,3}^2(n\Lambda) = \frac{g_{7,3}^2(\Lambda)}{n^3}=\frac{2\pi R_7}{2}\frac{g_{6,3}^2(\Lambda)}{n^3},
\end{equation}
where we assumed dimensionless bulk gauge coupling
$\hat{g}_{7,3}^2$ is on the UVFP: $g_{7,3}^2(\mu) = \hat{g}_{7,3}^2(\mu)/\mu^3 = \hat{g}_{7*,3}^2/\mu^3$, since 
the dimensionless bulk QCD coupling approaches very quickly to the UVFP value~\cite{HTY:2000uk}
(it is also true in the case without quark contributions). 
Then the induced four-fermion coupling is given by
\begin{equation}
\begin{split}
 \frac{G^{(7)}_6}{2 N_c}&= \frac{g^2_{6,3}(\Lambda)}{2 \Lambda^2}\cdot\frac{1}{2}\cdot\left[\frac{3}{4}
 \sum_{n = 1}^{n_{\rm KK}} \frac{\cos^2({n x_{70}}/R_7)}{n^5} \right]
 \\ 
 &=
\frac{g^2_{6,3}(\Lambda)}{2 \Lambda^2} \cdot\frac{1}{2}\cdot c_6^{(7)}(x_{70}) ,
\end{split} \label{eq:6dlag-GNJL07}
\end{equation} 
where $c_6^{(7)}(x_{70})$ is the order $O(1)$ coefficient including factors from Fierz transformation, 
the running effects of gauge coupling and the number of KK-modes.

It is convenient to use the dimensionless four-fermion coupling 
\begin{equation}
  g_D \equiv 2^{D/2} G_D \Lambda^{D-2} \NDA .
\label{dless-4coupling0}
\end{equation}
In our case $D=6$, 
the dimensionless induced four-fermion coupling from the 7-dimensional bulk
gluons is
\begin{equation}
g_6^{(7)}=2^3 G_6^{(7)} \Lambda^4 \NDA=c_6^{(7)}(x_{70}) \cdot 2^2 N_c \cdot
\hat{g}^2_{6,3}(\Lambda)\NDA =c_6^{(7)}(x_{70})\cdot\frac{3 N_c}{11} ,
\end{equation}
where 
we noted that the dimensionless gauge coupling $\hat{g}_{6,3}(\Lambda)^2$ defined by Eq.(\ref{dimlessgauge}) 
may be evaluated by the value at the UVFP; $\hat{g}_{6,3}(\Lambda)^2 \NDA
\simeq \hat{g}_{6*,3}\NDA=3/44$ 
as given in Eq.~(\ref{UVFP}).
As to the evaluation of $c_6^{(7)}$, 
sum of an infinite tower of KK modes in this case happen to be 
 given explicitly  by a finite 
number, although we implicitly assume 
that a bulk gauge theory is a cutoff theory. Then we can estimate  
the upper bound of $c_6^{(7)}$ exactly,
\begin{equation}
\begin{split}
 c_6^{(7)}(x_{70}) &=\frac{3}{4} \sum_{n = 1}^{n_{\rm KK}} 
  \frac{1}{n^5} \cos^2\frac{n x_{70}}{\Lambda^{-1}} \\
         &< c_6^{(7)}(x_{70}=0) < 
         \frac{3}{4} \sum_{n}^{\infty} \frac{1}{n^5} = \frac{3}{4} \zeta(5),
\end{split}
\end{equation}
hence the upper bound of $g_6^{(7)}$ ($N_c=3$) is given by 
\begin{equation}
 g_6^{(7)}=\frac{3 N_c}{11} c_6^{(7)}(x_{70}) < \frac{27}{44} \zeta(5) = 0.636297,
 \label{D=7g}
\end{equation}
which would have a chance to fulfill the condition of the condensation 
$g_6^{(7)} >g_6^{\rm crit}$, where $g_6^{\rm crit}$ ($1/2 < g_6^{\rm crit}< 2$) 
 is the value (depending on the gauge binding strength $\kappa$) on the critical line of the 
 6-dimensional
gauged NJL model~\cite{Gusynin:2004jp} (see later discussion in Sec. \ref{gnjl in 6d}).
\footnote{
If one exploited the compactification on $S^1$ instead of $S^1/Z_2$, the value of
$g_6^{(7)}$ would be twice larger than Eq.(\ref{D=7g}). However,
we actually need a freedom to tune the brane position,
as we shall discuss later, so that $S^1/Z_2$ is really needed. There is also an expectation~\cite{Dobrescu:1998dg}
 that $D>4$ bulk gluons would induce 
strong enough four-fermion interactions  to trigger the condensate even
if the top quark is fixed just on the 3-brane, 
$g_4^{(D)} >g_4^{\rm crit}$, where $g_4^{\rm crit}\simeq 1$ (the four-dimensional QCD coupling is small, i.e., $\kappa_3 \ll 1$ and hence $g_4^{\rm crit}$ is at the edge of the 
critical line of the four-dimensional gauge NJL model).
However, this expectation is also based on the compactification $T^{\delta}$ 
(and $\delta=D-4=4$) instead of $T^{\delta}/Z_2^{\delta/2}$. 
If we take $T^{\delta}/Z_2^{\delta/2}$ compactification, for the reason mentioned above,
we find that, as shown in appendix~\ref{app1}, 
$ g_4^{(D)} < g_4^{\rm crit}\simeq 1$
even for $D=8$. Thus the top fixed on the 3-brane actually does not condense,
besides the problem that such a scenario does not have seesaw partners of the top
which naturally arise as the KK modes when the top feels extra dimensions.
}
However, we shall later show that actual possible $g_6^{\rm crit}$ required for the top condensate {\it without tau 
condensate} (tMAC scale condition) is  $g_6^{\rm crit} >1.104$ (see Eq.(\ref{notautcond2})), which is not satisfied even by the
upper limit in Eq.(\ref{D=7g}). Then the $D=7$ bulk gluons are not enough for producing the four-fermion 
interaction strong enough to trigger the top condensation while forbidding the tau condensate.

By this point we may remark that if we estimate the sum only by the lowest KK-mode or only up till
the 4-th KK mode, $c_6^{(7)}(0)$ is given by
\begin{eqnarray}
 c_{6,\rm lowest}^{(7)}(0) &=& \frac{3}{4} , \label{c7-lowest}\\
 c_{6,\rm 4-th}^{(7)}(0) &=& 0.777256 \sim \frac{3}{4}\zeta(5) = 0.777696 \label{c7-4th}.
\end{eqnarray}
Hence, the summation of all the KK-mode effects is nearly equal to the summation only up till the 4-th effects.

Let us now consider the case of the gluons in the eight-dimensional bulk.
As in the above derivation, after the seventh and eighth dimensions are compactified ($D=8 \rightarrow D=6$)
 on $T^2/Z_2$, 
four-fermion interactions on the 5-brane are induced by the gluon Kaluza-Klein(KK) modes exchange.

The Lagrangian reads:
\begin{equation}
\mathcal{L}_{8D} = \delta(x_7 - x_{70})\delta(x_8 - x_{80}) \bar{\psi}(X) i \Gamma^{M} D_{M} \psi(X),
\label{eq:8dlag-1}
\end{equation}
where
\begin{equation*}
 D_{M} = \partial_{M} - i g_{8,3} G^a_{M}(x , y , z) T^a,
\end{equation*}
with $x=x_{\mu}, y=x_{5,6}, z=x_{7,8}, X=x,y$.
This Lagrangian is gauge invariant on the 5-brane$(M = \mu,5,6)$.
In order to compactify the seventh and eighth dimensions, we impose the boundary conditions ($T^2/Z_2$-compactification): 
\begin{eqnarray}
 G_M(X,x_7,x_8) &=& G_M(X,x_7+2 \pi R_7,x_8) \label{eq:bc1}\\
 &=&G_M(X,x_7,x_8 + 2 \pi R_8), \notag \\
 G_M(X,x_7,x_8) &=& G_M(X,-x_7,-x_8). \label{eq:bc2} 
\end{eqnarray}

Hence the bulk gluons $G_M$ are decomposed as follows ($R_7=R_8=\Li$).
\begin{equation}
\begin{split}
 G_M (X , x_7 , x_8) = \frac{1}{2 \pi \Li} \Biggl[ G_{M,00}(X) &+ \sqrt{2} \sum_{n = 1}^{n_{\rm KK}} G^{[n]}_{M,c0}(X) 
\cos\frac{n x_7}{\Li}\\
&+ \sqrt{2} \sum_{n = 1}^{n_{\rm KK}} G^{[n]}_{M,0c}(X) \cos\frac{n x_8}{\Li}\\
&+ 2 \sum_{n_1,n_2 = 1}^{n_{\rm KK}} G^{[n_1,n_2]}_{M,cc}(X) \cos\frac{n_1 x_7}{\Li} \cos\frac{n_2 x_8}{\Li} \\
&+ 2 \sum_{n_1,n_2 = 1}^{n_{\rm KK}} G^{[n_1,n_2]}_{M,ss}(X) \sin\frac{n_1 x_7}{\Li} \sin\frac{n_2 x_8}{\Li} \Biggr].
\end{split} \label{eq:8d-gkk1}
\end{equation}

From the interaction between the fermions on the 5-brane and $G_M$, 
we derive four-fermion interactions on the 5-brane via one-coloron exchange. 
In consequence, we have 
\begin{equation}
\mathcal{L}_{6D} =\bar{\psi}(X) i \Gamma^{M} D_{M,00} \psi(X) + \mathcal{L}_{\rm 4F},
\label{eq:6dlag-GNJL1}
\end{equation}
with
\begin{equation}
D_{M,00}= \partial_{M} - i g_{6,3} G^a_{M,00}(X) T^a, \quad g_{6,3}^2=2\frac{g^2_{8,3}}{(2 \pi \Lambda^{-1})^2}
\Bigg|_{\mu=\Lambda}
\end{equation} 
and
\begin{equation}
\begin{split}
 \mathcal{L}_{\rm 4F} = -\frac{2}{(2 \pi \Li)^2}
 &\Bigl[\sum_{n = 1}^{n_{\rm KK}} g^2_{8,3} \frac{\cos^2({n x_{70}}/\Li)}{2 M^2_n}
 +\sum_{n = 1}^{n_{\rm KK}} g^2_{8,3} \frac{\cos^2({n x_{80}}/\Li)}{2 M^2_n} \\
 &+2\sum_{n_1,n_2 = 1}^{n_{\rm KK}} g^2_{8,3} \frac{\cos^2({n_1 x_{70}}/\Li) \cos^2({n_2 x_{80}}/\Li)}{2 M^2_{\vec{n}}} \\
 &+2\sum_{n_1,n_2 = 1}^{n_{\rm KK}} g^2_{8,3} \frac{\sin^2({n_1 x_{70}}/\Li) \sin^2({n_2 x_{80}}/\Li)}{2 M^2_{\vec{n}}} \Bigr]
 \times (\bar{\psi} \Gamma^M T^a \psi)(\bar{\psi}\Gamma_M T^a \psi).
\end{split} \label{eq:6dlag-GNJL2}
\end{equation}

Furthermore, after the Fierz transformations, scalar and pseudoscalar channels in (\ref{eq:6dlag-GNJL2}) are
\begin{eqnarray}
 \mathcal{L}_{\rm 4F}^\prime &=& \frac{3}{4} \cdot \frac{1}{2} \cdot \frac{2}{(2 \pi \Li)^2} 
 \Bigl[{\sum_{n = 1}^{n_{\rm KK}} g^2_{8,3} \frac{\cos^2({n x_{70}}/\Li)}{2 M^2_n}}
 +{\sum_{n = 1}^{n_{\rm KK}} g^2_{8,3} \frac{\cos^2({n x_{80}}/\Li)}{2 M^2_n}} \notag \\
 &&\hspace*{2cm} +2{\sum_{n_1,n_2 = 1}^{n_{\rm KK}} g^2_{8,3} \frac{\cos^2({n_1 x_{70}}/\Li) \cos^2({n_2 x_{80}}/\Li)}{2 M^2_{\vec{n}}}} \notag \label{eq:6dlag-GNJL3} \\
 &&\hspace*{2cm}+2{\sum_{n_1,n_2 = 1}^{n_{\rm KK}} g^2_{8,3} \frac{\sin^2({n_1 x_{70}}/\Li) \sin^2({n_2 x_{80}}/\Li)}{2 M^2_{\vec{n}}}} \Bigr] 
 \times \Bigl[(\bar{\psi} \psi)^2 + (\bar{\psi} i \Gamma_A \psi)^2 \Bigr] , \\
 &=& \frac{G_6^{(8)}}{2N_c} \Bigl[(\bar{\psi} \psi)^2 + (\bar{\psi} i \Gamma_A \psi)^2 \Bigr],
\end{eqnarray}
where $G^{(8)}_6$ is the dimensionful four-fermion coupling
on the 5-brane
and $\vec{n}=(n_1,n_2)$.

The coefficient of the first term in the brackets of (\ref{eq:6dlag-GNJL3}) becomes
\begin{equation}
\begin{split}
 \sum_{n = 1}^{n_{\rm KK}} g^2_{8,3} \frac{\cos^2({n x_{70}}/\Li)}{2 M^2_n} &= 
 \sum_{n=1}^{n_{\rm KK}} g^2_{8,3}(n\Lambda) \frac{\cos^2(nx_{70}/\Li)}{2 (n\Lambda)^2} \\
 &= \frac{g^2_{8,3}(\Lambda)}{2 \Lambda^2} \sum_{n = 1}^{n_{\rm KK}} \frac{\cos^2({n x_{70}}/\Li)}{n^6}\, ,
\end{split} \label{eq:6dlag-GNJL4}
\end{equation}
where we have used again the fact that dimensionless bulk gauge coupling $\hat{g}^2_{8,3}$ is approximately near the UVFP
and set 
\begin{equation}
 g_{8,3}^2(n\Lambda)=\frac{g_{8,3}^2(\Lambda)}{n^4}.
 \label{runningg8}
\end{equation}

In the same way, the second and third terms in the brackets of  (\ref{eq:6dlag-GNJL3}) become 
\begin{equation}
 {\sum_{n = 1}^{n_{\rm KK}} g^2_{8,3} \frac{\cos^2({n x_{80}}/\Li)}{2 M^2_n}} =
\frac{g^2_{8,3}(\Lambda)}{2 \Lambda^2} \sum_{n = 1}^{n_{\rm KK}} \frac{\cos^2({n x_{80}}/\Li)}{n^6}, \label{eq:6dlag-GNJL5}
\end{equation}
\begin{equation}
\begin{split}
 \sum_{n_1,n_2 = 1}^{n_{\rm KK}} g^2_{8,3} & \frac{\cos^2 ({n_1 x_{70}}/\Li) \cos^2({n_2 x_{80}}/\Li)}{2 M^2_{\vec{n}}} \\  
 &= \frac{g^2_{8,3}(\Lambda)}{2 \Lambda^2} \sum_{n_1, n_2} \frac{1}{(n_1^2 + n_2^2)^3} 
 \cos^2\frac{n_1 x_{70}}{\Lambda^{-1}} \cos^2\frac{n_2 x_{80}}{\Lambda^{-1}},
\end{split} \label{eq:6dlag-GNJL6}
\end{equation}
\begin{equation}
\begin{split}
 \sum_{n_1,n_2 = 1}^{n_{\rm KK}} g^2_{8,3} & \frac{\sin^2 ({n_1 x_{70}}/\Li) \sin^2({n_2 x_{80}}/\Li)}{2 M^2_{\vec{n}}} \\  
 &= \frac{g^2_{8,3}(\Lambda)}{2 \Lambda^2} \sum_{n_1, n_2} \frac{1}{(n_1^2 + n_2^2)^3} 
 \sin^2\frac{n_1 x_{70}}{\Lambda^{-1}} \sin^2\frac{n_2 x_{80}}{\Lambda^{-1}}.
\end{split} \label{eq:6dlag-GNJL6-2}
\end{equation}

Since we define the six-dimensional gauge coupling as $g^2_{6,3}(\Lambda) \equiv 2 g^2_{8,3}(\Lambda)/{(2 \pi \Li)^2}$, 
the total coefficient of 
four-fermion operator in (\ref{eq:6dlag-GNJL3}) is given by
\begin{equation}
\begin{split}
 \frac{G^{(8)}_6}{2 N_c}
&= \frac{g^2_{6,3}(\Lambda)}{2 \Lambda^2} \cdot \frac{1}{2} \cdot \frac{3}{4} \Bigg[ \sum_{n_1 = 1}^{n_{\rm KK}} \frac{\cos^2({x_{70}}/\Li)}{n^6_1} 
 + \sum_{n_2 = 1}^{n_{\rm KK}} \frac{\cos^2({x_{80}}/\Li)}{n^6_2} \\ 
&\hspace*{2cm}+ \sum_{n_1, n_2} \frac{2}{(n_1^2 + n_2^2)^3} 
 \Bigl( \cos^2\frac{n_1 x_{70}}{\Lambda^{-1}} \cos^2\frac{n_2 x_{80}}{\Lambda^{-1}} 
+ \sin^2\frac{n_1 x_{70}}{\Lambda^{-1}} \sin^2\frac{n_2 x_{80}}{\Lambda^{-1}} \Bigr) 
 \Bigg]. \\
 &=\frac{g^2_{6,3}(\Lambda)}{2 \Lambda^2}\cdot \frac{1}{2}\cdot c_6^{(8)}(x_{70},x_{80}), \\
\end{split} \label{eq:6dlag-GNJL7}
\end{equation}
where  $c_6^{(8)}(x_{70},x_{80})$ is the order $O(1)$ coefficient including factors from Fierz transformation, 
the running effects of gauge coupling and the number of KK-modes.
Hence, the dimensionless induced four-fermion coupling  defined in Eq.(\ref{dless-4coupling0}) 
 is given by
\begin{equation}
 g_6^{(8)}  \equiv 2^{3} G_6^{(8)} \Lambda^{4} \NDA 
 =c_6^{(8)}(x_{70},x_{80})\cdot 2^2 N_c \cdot \hat{g}_{6,3}^2\NDA 
= c_6^{(8)}(x_{70},x_{80})\cdot\frac{3 N_c}{11} ,
\label{6dim4fermi8dimgluons}
\end{equation}
where we again used the fact that the dimensionless gauge coupling 
on the 5-brane, $\hat{g}_{6,3}^2\NDA$, (Eq.(\ref{dimlessgauge})) 
is approximately the value on the 
UVFP: $\hat{g}_{6,3}^2\NDA=\hat{g}_{6*,3}^2\NDA=3/44$.

We now evaluate the upper bound of  $g_6^{(8)}$ given at $x_{70}=x_{80}=0$.
 From the experience of $D=7$ case (see Eqs. (\ref{c7-lowest}) (\ref{c7-4th})), we may expect the summation is approximately saturated only by
the lowest KK-mode or at most by the summation till the 4-th KK mode in (\ref{eq:6dlag-GNJL7}):
\begin{eqnarray}
 c^{(8)}_{6,\rm lowest}(0,0) &=& \frac{3}{4} \cdot \frac{2}{1^6} = 1.5 \\
 c^{(8)}_{6,\rm 4-th} (0,0)&=& \frac{3}{4} \cdot \Bigl[\frac{2}{1^6} + \frac{2}{2^3} + \frac{2}{2^6} + \frac{4}{5^3} \Bigr] \sim 1.73 .
\end{eqnarray}
Actually, we show in Appendix~\ref{app2} that the actual value of the summation is numerically almost saturated by the sum 
only till the 4-th KK modes, if we assume that the {\it dimensionless}
gauge coupling  between the fermions on the brane and the n-th
KK-mode of the $8$-dimensional bulk gauge field is equal for each KK mode, i.e.,
$\hat{g}_{8,3} (n\Lambda)=\hat{g}_{8*,3}$ or Eq.(\ref{runningg8}). 
On the other hand, if we literally did sum an infinite number
of KK modes contributions (assuming the bulk theory is well-defined without ultraviolet 
cutoff), 
we would get a divergent result in contrast to the case of $D=7$
(one extra dimension case). Moreover, there is a large anomalous 
dimension for the four-fermion operators~\cite{HTY:2000uk,Gusynin:2004jp} which may prevent the naive dimensional
suppression of the four-fermion operators induced by the higher KK modes.  However, it was pointed out~\cite{Bando:1999}
that considering the recoil effect of the brane, the gauge coupling is suppressed exponentially
\begin{equation}
 \hat{g}_{D,3} (n\Lambda) \sim 
 \exp(-n^2/R_D^2),
 \label{gauge-sup}
\end{equation}
where
$R_D$ is the compactified radii of the $D$-dimensions. Due to such an exponential decreasing,
the actual summation of KK-mode effects will converge even if we assumed the bulk theory
without ultraviolet cutoff, and hence we expect that the actual sum  
may be even nearly equivalent to the lowest KK-mode only or at most up till the 4-th KK modes.

Then we have (for $N_c=3$)
\begin{eqnarray}
 g_6^{(8)} &<& \frac{3 N_c}{11} \cdot c^{(8)}_{6,\rm lowest}(0,0) \sim 1.22 \hspace*{5mm} (\text{by the lowest KK-mode}) ,\label{lowest}\\
 g_6^{(8)} &<& \frac{3 N_c}{11} \cdot c^{(8)}_{6,\rm 4-th} (0,0) \sim 1.42 \hspace*{6mm} (\text{by the 4-th KK-modes}) ,\label{4th}
\end{eqnarray}
which is compared with the result of the $D=7$ bulk in Eq.(\ref{D=7g}),  $g_6^{(7)} < 0.636297$.
Then the $D=8$ bulk gluons can induce a strong enough four-fermion interaction to trigger  the top condensate without tau 
condensate (tMAC scale condition)  $g_6^{(8)}  > g_6^{\rm crit} $ where $g_6^{\rm crit}$ will be shown later to satisfy 
$g_6^{\rm crit}>1.104$, the condition of no tau condensation.
coming from the brane position dependence 
in Eq. (\ref{eq:6dlag-GNJL7}) :
$0 <g_6^{(8)}$ and $0.18 <g_6^{(8)}$ for the sum till the lowest and the 4-th KK modes,
respectively.
Hence 
we conclude that the allowed regions for the summation by the lowest or the 4-th KK-modes effects are
\begin{eqnarray}
 0<g_6^{(8)}<1.22 & \hspace*{2mm} & (\text{by the lowest KK-mode}), \label{lowest-g} \\
 0.18<g_6^{(8)}<1.42 & \hspace*{2mm} & (\text{by the 4-th KK-modes}). \label{4th-g}
\end{eqnarray}

\section{\lowercase{t}MAC Scale in the 6-Dimensional Gauged NJL Model with the Induced 
Four-Fermion Interaction}
\label{gnjl in 6d}

First, we briefly depict the $D(=6)$-dimensional gauged NJL dynamics following Ref.~\cite{Gusynin:2004jp} which is based on the improved ladder SD equation with the argument of the
running (dimensionful) gauge coupling identified with the gauge boson momentum. 
The D-dimensional fermion propagator takes the form $ i S^{-1}(p) = A(-p^2)\, [\fsl{p} - \Sigma(-p^2)]$. With the above momentum identification we take a particular gauge
(''nonlocal gauge'') in order to keep $A(-p^2) \equiv 1$. Then,
the SD equation becomes a gap equation for  
the  dynamical mass function $\Sigma (x\equiv -p^2)$:
\begin{eqnarray}
 \Sigma(x)
      &=& \sigma + (D-1+ \xi )
          \kappa_D \int_0^{\Lambda^2}\!\!\! dy 
          \frac{y^{D/2-1}\Sigma(y)}{y + \Sigma^2(y)} \nonumber \\
      &&    \qquad \times \frac{1}{[\max(x,y)]^{D/2-1}},
      \label{imp_ladder_SD}
\end{eqnarray}
where $\sigma$ is 
\begin{equation}
  \sigma 
         = \frac{g_D}{\Lambda^{D-2}} \int_0^{\Lambda^2}
           dx x^{D/2-1} \dfrac{\Sigma(x)}{x + \Sigma^2(x)},
  \label{eq:vev_sigma}
\end{equation}
and  $\xi$ is the gauge fixing parameter which is taken to be 
$\xi=-(D-1)(D-4)/D$ (``nonlocal gauge''), and
we have assumed that  
the binding strength $\kappa_D(\mu)$  is almost constant
 over the entire energy region relevant to the 
SD equation. 

\begin{figure}
\begin{tabular}{cc}
\begin{minipage}{0.5\hsize}
\begin{flushleft}
  \hspace*{1.2cm} {(a) \hspace*{5mm} \underline{$\delta=1\,,R_6^{-1}=10\,\TeV$}}
\end{flushleft}
  \vspace*{-0.9cm}
\includegraphics[width=0.75\hsize, clip]{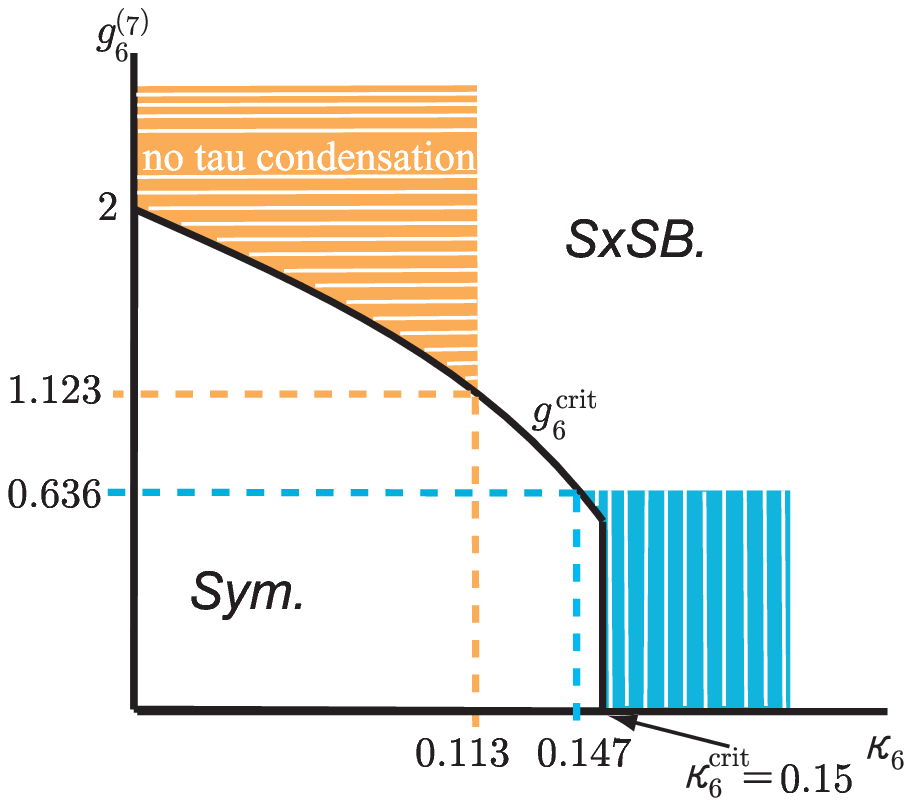}
\end{minipage}
\begin{minipage}{0.5\hsize}
\begin{flushleft}
  \hspace*{1.2cm} {(b) \hspace*{5mm} \underline{$\delta=1\,,R_6^{-1}=1\,\TeV$}}
\end{flushleft}
  \vspace*{-0.9cm}
\includegraphics[width=0.75\hsize, clip]{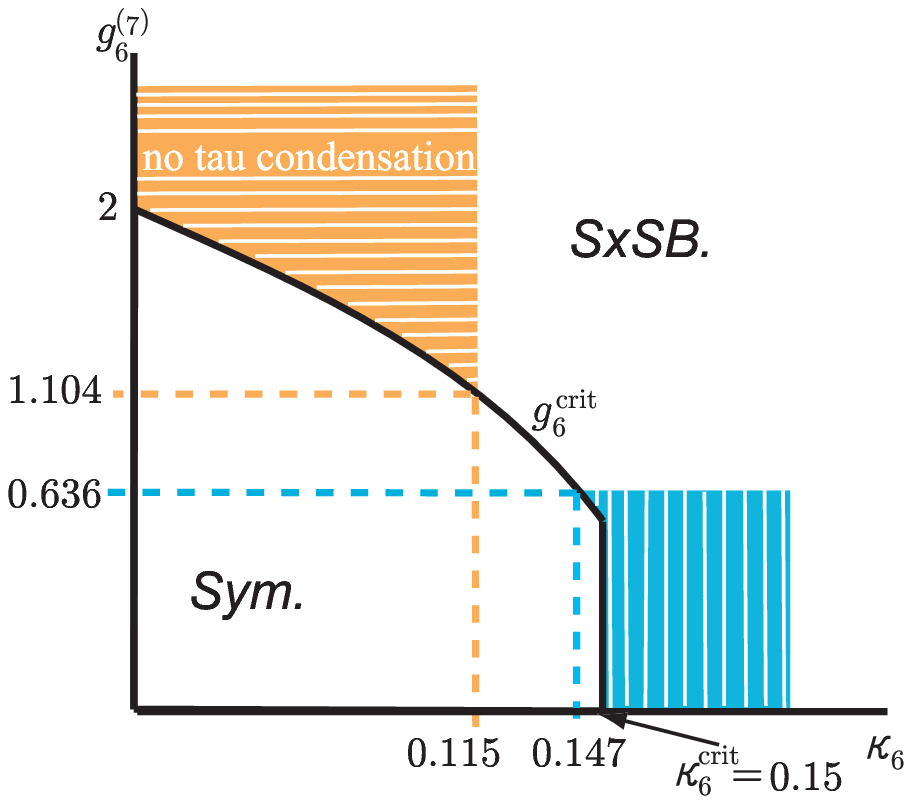}
\end{minipage}
\end{tabular}
\caption{The phase diagram of D(=6)-dimensional gauged NJL model~\cite{Gusynin:2004jp}
with the induced four-fermion coupling by the
7-dimensional bulk gluons (one compactified dimension $\delta=1$).
 The critical line in Eq.(\ref{criticalline1}) is denoted by 
$g_6^{\rm crit}$ with the nonlocal gauge estimation. 
The region above $g_6^{\rm crit}$ is the $S\chi SB$-phase, and that below 
is  the $Symmetric$-phase. 
The vertical stripe pattern regions are from Eq.~(\ref{induced7}): 
$g_6^{\rm crit}(\kappa_t) <g_6^{\rm induced} <0.636$ or equivalently
 $\kappa_t > \kappa^{\rm crit} >\kappa_6^{\rm crit}(0.636)=0.147$.  The 
horizontal stripe pattern regions are those satisfying 
         Eq.(\ref{notautcond}), namely the region for the top condensate 
         without tau condensation:
 (a) is for $R_6^{-1}=10\,\TeV$ with $(\kappa_t(\Lambda_{\tau}),\, g_6^{\rm crit}(\kappa_t(\Lambda_{\tau}))) =(0.113, \, 1.123)$ 
 and (b) is for $R_6^{-1}=1\,\TeV$ with $(\kappa_t(\Lambda_{\tau}),\, g_6^{\rm crit}(\kappa_t(\Lambda_{\tau}))) =(0.115, \, 1.104)$. The tMAC scale satisfying
 Eq.(\ref{tMAC1}) would be the overlap
 region between regions of the vertical and the horizontal stripe pattern, which does not
 exist in either case (a) or (b).
         }
\label{7-6-crit}
\end{figure}

Solving the SD equation, we find the critical line in $(\kappa_{D}, g_D)$-plane 
separating the broken phase $\Sigma \ne 0$ and unbroken phase $\Sigma =0$, which
takes the form:
\begin{equation}
g_D
= \frac{\frac{D}{2} -1}{4}
 \left( 1+ \sqrt{1-\kappa_D/\kappa_D^{\rm crit} } \right)^2\, ,
 \label{criticalline1}
\end{equation}
for $0< \kappa_D < \kappa_D^{\rm crit}$, or $\frac{1}{4}(\frac{D}{2}-1) <g_D < \frac{D}{2}-1$
(Fig.~\ref{7-6-crit}(a) for
$D=6$),
where 
$\kappa_D^{\rm crit}$ is the critical binding strength of gauge interactions
which was obtained from the SD equation 
without four-fermion 
coupling $g_D =0$ or $\sigma =0$ in the nonlocal gauge as 
given in Eq.(\ref{criticalkappa})~\cite{Gusynin:2002cu}:   
\begin{equation}
\kappa_D^{\rm crit}
    = \dfrac{D}{32} \dfrac{D-2}{D-1} \quad =0.15 \,\,(D=6)
    \hspace*{2mm} (\text{nonlocal gauge fixing}) .
    \label{critkappanonlocal}
\end{equation}

From our consideration in Sec.~\ref{model},
there are induced four-fermion interactions for the top and the bottom but not for the tau.
Hence, while the critical binding strength of the tau remains the same as that in 
Eq.(\ref{critkappanonlocal}), $\kappa_6^{\rm crit} =0.15$, 
that of the top and the bottom channels 
decreases, for $g_D=g_6^{\rm induced} >1/2$, down to 
that of the gauged NJL model, $\kappa_6^{\rm crit}\rightarrow \kappa^{\rm crit} (<\kappa_6^{\rm crit})$,
where $\kappa^{\rm crit}$
is the 
critical $\kappa_6$ value for the 
top and the bottom in the presence of the induced four-fermion interaction and is given through
 the inversion, $\kappa_D= \kappa_D(g_D)$, 
of the critical line  Eq.(\ref{criticalline1}) (for $D=6$)  as
\begin{eqnarray}
 && \kappa^{\rm crit}\equiv \kappa_D (g_D) \Big|_{g_D=g_6^{\rm induced}} \quad (D=6),
\label{critkappa}\\
&&g_6^{\rm induced} = g_6^{(7)}, g_6^{(8)}
\end{eqnarray}
with
$g_6^{(7)}$ and $g_6^{(8)}$ being given by Eq.(\ref{D=7g}) and Eqs. (\ref{lowest-g}), (\ref{4th-g}),
respectively.
Because of this lowering the
critical binding strength for the top/bottom, we expect 
that 
the top condensation becomes possible even if $\kappa_t <\kappa_6^{\rm crit}$.

If, instead of nonlocal gauge value in Eq.(\ref{critkappanonlocal}), we may 
take the Landau gauge fixing
 which  makes $A(-p^2)=1$ for a 
different choice of momentum identification for the
scale of the
running coupling in the SD equation.  Then we would 
have  $\kappa_D^{\rm crit} = \frac{1}{8}\frac{D-2}{D-1}$  
somewhat smaller than that of the nonlocal gauge~\cite{HTY:2000uk}. 
However, the SD equation in the Landau
gauge for such a momentum identification
is not consistent with the axialvector Ward-Takahashi identity~\cite{Kugo:1992pr}.
So, throughout this paper we adopt the nonlocal gauge fixing. If we find a condensate in the nonlocal gauge, then there always exists a condensate for the Landau gauge. Thus our conclusion of
the existence of a condensate is fairly independent of this ambiguity of the SD equation in contrast to
the tMAC analysis of Ref.~\cite{HTY2003} where the existence of a tMAC scale
for the $D=8$ model critically depends on the usage of the Landau gauge value. 
There are actually some other ambiguities about the estimate of $\kappa_6^{\rm crit}$~\cite{HTY2003}
and hence $\kappa^{\rm crit}$ as well:
\begin{enumerate}
\item The non-ladder corrections to the SD equation, which is known~\cite{Appelquist:1988yc} in the 4-dimensional walking technicolor to decrease $\kappa_6^{\rm crit}$ down to 1-20 \%.
\item  Finite size effects of the $R_5^{-1}=R_6^{-1}$ compared with $\Lambda=R_7^{-1}=R_8^{-1}$
in the SD equation, which would increase $\kappa_6^{\rm crit}$ and $\kappa^{\rm crit}$.
\item The running 
effects of $\kappa_D(\mu)$ in the SD equation, 
which would also increase $\kappa_6^{\rm crit}$ and $\kappa^{\rm crit}$.

\item There is also a scheme-dependence of binding strength's $\kappa_\alpha(\mu)$: In our estimate we used $\overline{\rm MS}$ scheme for the bulk gauge couplings and hence 
$\kappa_\alpha(\mu)$. In Ref.~\cite{HTY2003} the results were compared with 
the proper-time regularization scheme and the scheme-dependence was found to be small. 
\end{enumerate}
Understanding all these ambiguities which could change the estimate in opposite directions, 
we shall use the value of Eq.(\ref{critkappanonlocal}) and Eq.(\ref{criticalline1}) 
as a reference value with 
possible errors at most 20 \%.

\begin{figure}
\begin{tabular}{cc}
\begin{minipage}{0.5\hsize}
\begin{flushleft}
  \hspace*{1.2cm} {(a)}
\end{flushleft}
  \vspace*{-0.9cm}
\includegraphics[width=0.75\hsize, clip]{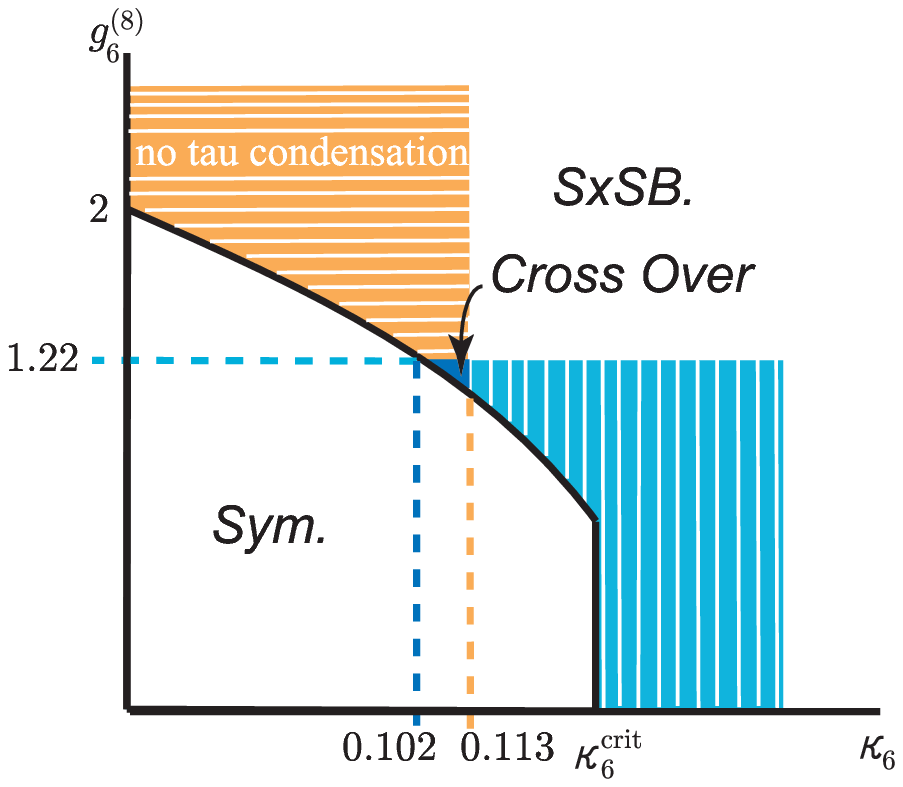}
\end{minipage}
\begin{minipage}{0.5\hsize}
\begin{flushleft}
    \hspace*{1.2cm} {(b)}
  \end{flushleft}
  \vspace*{-0.9cm}
\includegraphics[width=0.75\hsize, clip]{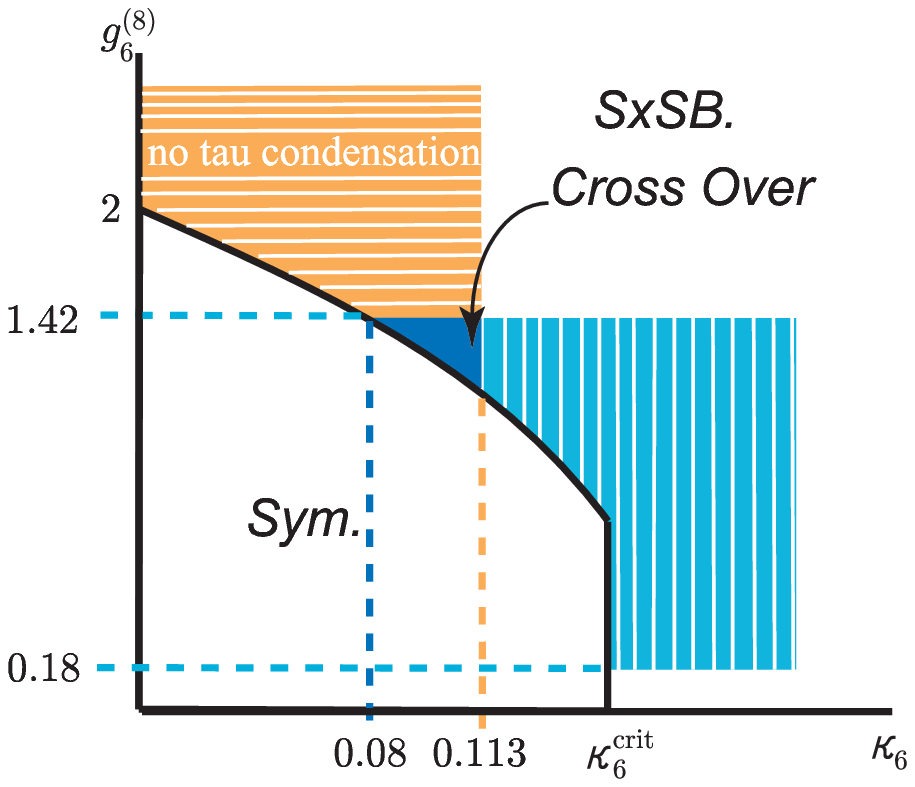}
\end{minipage} 
\end{tabular}
\caption{The phase diagram and the tMAC region by the nonlocal gauge fixing method;
         These figures are for $R_6^{-1}=10\,\TeV $case.
         (a) is estimated by the lowest KK-mode only, and 
         (b) is by the summation till the 4-th KK-modes.
         The vertical dashed lines are the value of $\kappa_t$ 
         at scale $\Lambda_{\tau}$.
         The vertical stripe pattern regions are the allowed regions from Eq.(\ref{lowest-g}), (\ref{4th-g}).
         In order to make top quark condense, these vertical stripe region and the horizontal stripe
         (no tau condensation) regions 
         must have a cross over regions.
         In this figure, there exist cross over regions (shaded regions).
         }
\label{crit-MAC}
\end{figure}

Now we discuss the existence of the tMAC scale in our model, namely the scale where
only the top condenses while other fermions do not.  
We look for the  tMAC scale $\mu=\Lambda_{\rm tM}=\Lambda=R_7^{-1}=R_8^{-1}$ 
such that
\begin{eqnarray}
&&\kappa_b(\Lambda_{\rm tM})< \kappa^{\rm crit} <
\kappa_t(\Lambda_{\rm tM}),
\label{tMAC00}\\
&& \kappa_\tau(\Lambda_{\rm tM}) < \kappa_6^{\rm crit} =0.15.
\label{tMAC01}
\end{eqnarray}
Note that Eq.(\ref{tMAC01}) is the condition that the tau condensation does not take place, the value of $\Lambda_{\rm tM}$ for which can be read off  from Fig.~\ref{fig:tMAC0} as
\begin{equation}
 \Lambda_{\rm tM} < \Lambda_{\tau} =
 \begin{cases}
 10.8 R_6^{-1}&(\text{for}\,R_6^{-1}=10\,\TeV) \\[3mm]
 11.0 R_6^{-1}&(\text{for}\,R_6^{-1}=1\,\TeV),
 \end{cases}
 \label{notaucond}
\end{equation}
where $\Lambda_\tau$ is defined by $\kappa_\tau(\Lambda_\tau) = \kappa_6^{\rm crit} =0.15$.
Then from Fig.\ref{fig:tMAC0} we further read the no tau condensation condition
in terms of $\kappa_t(\Lambda_{\rm tM})$ as:
\begin{equation}
  \kappa_t(\Lambda_{\rm tM}) < \kappa_t(\Lambda_{\tau})=
 \begin{cases}
\kappa_t(10.8 R_6^{-1}) = 0.113 &(\text{for}\,R_6^{-1}=10\TeV) \\[3mm]
\kappa_t(11.0 R_6^{-1}) =0.115 &(\text{for}\,R_6^{-1}=1\TeV),
 \end{cases}
 \label{notautcond}
\end{equation}
the region shown by the horizontal stripe pattern  in 
Figs.\ref{7-6-crit} and \ref{crit-MAC}.
Then the tMAC scale is a combination of Eq.(\ref{tMAC00}) and Eq.(\ref{notautcond}):
\begin{eqnarray}
\kappa_b(\Lambda_{\rm tM})< \kappa^{\rm crit} <
\kappa_t(\Lambda_{\rm tM})<\kappa_t(\Lambda_{\tau}).
\label{tMAC1}
\end{eqnarray}

Note that Eq.(\ref{notautcond})
is converted by the critical line Eq.(\ref{critkappanonlocal}) into 
\begin{eqnarray}
g_6^{\rm crit}(\kappa_t(\Lambda_{\rm tM}))>g_6^{\rm crit}(\kappa_t(\Lambda_\tau))
=
\begin{cases}
  1.123&(\text{for}\,R_6^{-1}=10\,\TeV) \\[3mm]
  1.104&(\text{for}\,R_6^{-1}=1\,\TeV),
 \end{cases}
\label{notautcond2}
\end{eqnarray}
where  $g_6^{\rm crit}(\kappa_t(\Lambda_{\rm tM}))$ is the critical $g_D (D=6)$ value for the top, which is given by the critical line  Eq.(\ref{criticalline1}) (for $D=6$) as $g_6^{\rm crit}(\kappa_t(\Lambda_{\rm tM}))=
g_D (\kappa_D=\kappa_t(\Lambda_{\rm tM}))$.
Then the tMAC scale $\Lambda_{\rm tM}$ may be defined in another equivalent way:
\begin{eqnarray}
   g_6^{\rm crit}(\kappa_t(\Lambda_\tau))< g_6^{\rm crit}(\kappa_t(\Lambda_{\rm tM})) <  g_6^{\rm induced}  <g_6^{\rm crit}(\kappa_b(\Lambda_{\rm tM})),
    \label{tMAC2}
\end{eqnarray}
where   $g_6^{\rm crit}(\kappa_b(\Lambda_{\rm tM}))$ is the critical $g_D (D=6)$ value for the  bottom defined  similarly to  $g_6^{\rm crit}(\kappa_t(\Lambda_{\rm tM}))$.
\\

\underline{$7D \to 6D$-case}\\
Let us first discuss that there is no tMAC scale in the case of the gluons in the 7-dimensional bulk.
In this case  
 the upper bound for $ g_6^{\rm induced}=g_6^{(7)}$ is given by Eq.~(\ref{D=7g}):
\begin{equation} 
 g_6^{\rm induced}= g_6^{(7)}<\frac{27}{44}\zeta(5)= 0.636,
 \label{induced7}
\end{equation}
which is much smaller than the value required by the condition of no tau condensation, 
Eq.(\ref{notautcond2}),
even when a possible ambiguity up to 20\%  in
 the estimate of the critical line by the 
SD equation is considered. Then there is no tMAC scale satisfying Eq.(\ref{tMAC2}).
Equivalently, Eq.(\ref{induced7})
implies that the top condensate would take place if 
 \begin{equation}
  \kappa_t(\Lambda_{\rm tM}) > \kappa^{\rm crit}\equiv  \kappa_6 (g_6)\Big|_{g_6= g_6^{\rm induced}} > 
  \kappa_6 (g_6)\Big|_{g_6=0.636} =0.147
  \label{topcondbyinduced}
 \end{equation}
(the vertical stripe pattern region in  Fig.~\ref{7-6-crit}),
which has no overlap with Eq.(\ref{notautcond}) 
 (horizontal stripe pattern
region in  Fig.~\ref{7-6-crit}).
Then there is obviously no tMAC scale satisfying Eq.(\ref{tMAC1}).
The induced four-fermion coupling $g_6^{(7)}$ is not strong enough, or
equivalently the reduction $\kappa_6^{\rm crit}\rightarrow \kappa^{\rm crit}$ 
is not enough in this case.\\
 
\newpage

\underline{$8D \to 6D$-case}\\

\begin{figure}
\begin{tabular}{cc}
\begin{minipage}{0.5\hsize}
\begin{flushleft}
  \hspace*{2mm} {(a) \hspace*{6mm} $R_6^{-1}=10\,\TeV$, lowest}
\end{flushleft}
  \vspace*{-1.2cm}
\begin{center}
 \includegraphics[width=\hsize, clip]{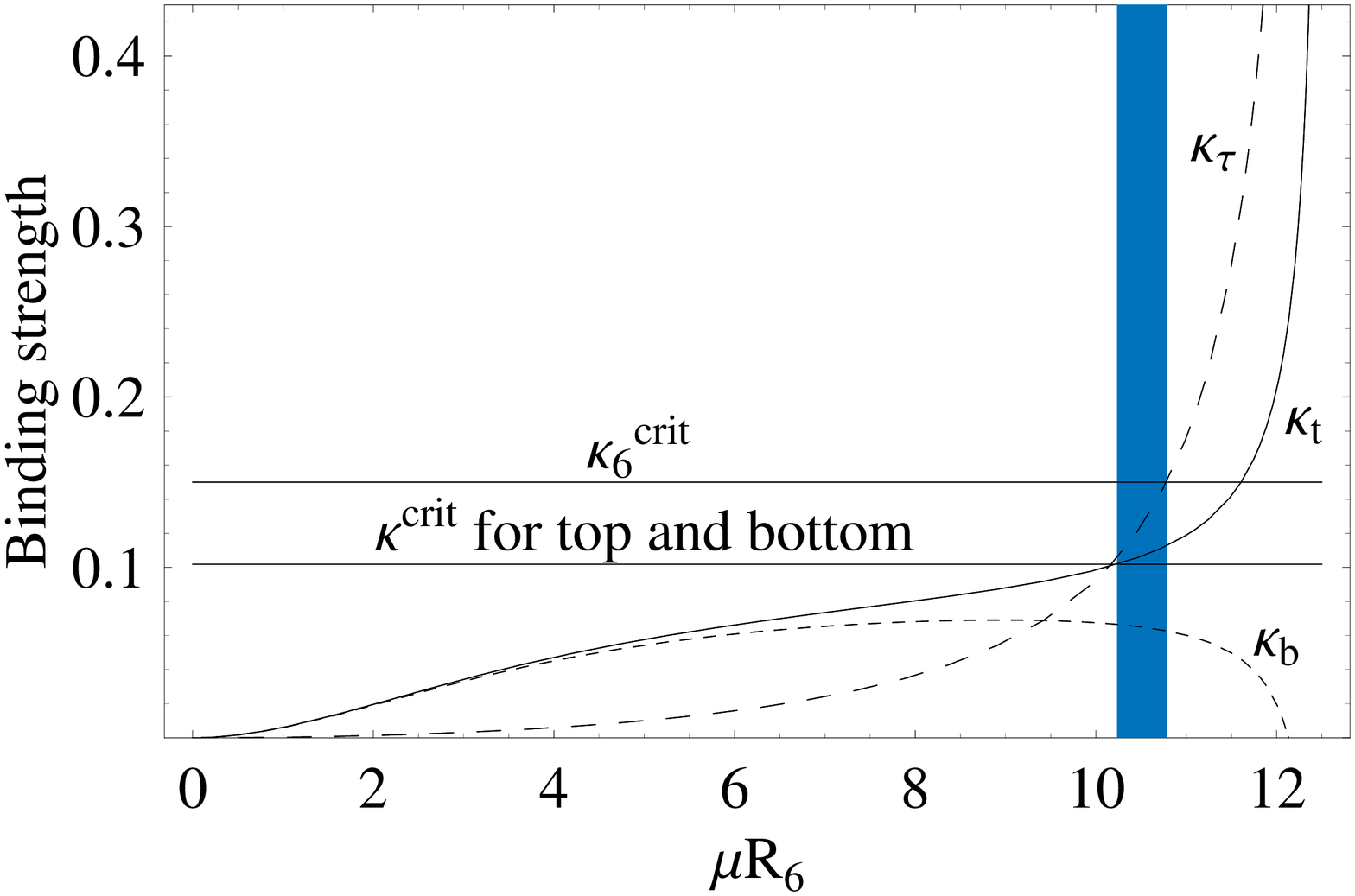}
\end{center}
\end{minipage}
\begin{minipage}{0.5\hsize}
 \begin{flushleft}
    \hspace*{1mm} {(b) \hspace*{6mm} $R_6^{-1}=10\,\TeV$, 4-th}
  \end{flushleft}
  \vspace*{-1.2cm}
\begin{center}
 \includegraphics[width=\hsize, clip]{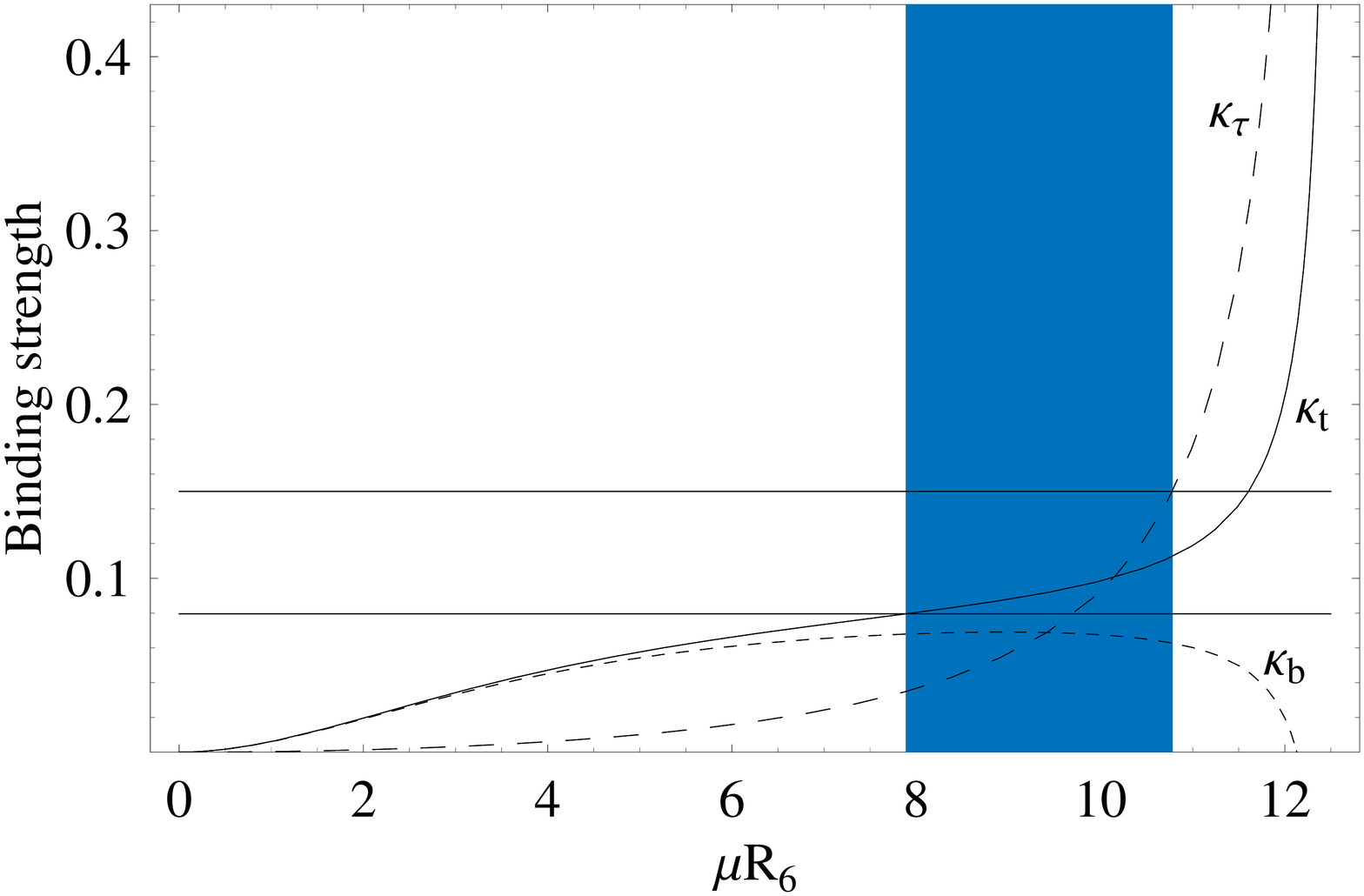}
\end{center}
\end{minipage} \\ \hspace*{2cm} \\
\begin{minipage}{0.5\hsize}
 \begin{flushleft}
    \hspace*{2mm} {(c) \hspace*{6mm} $R_6^{-1}=1\,\TeV$, lowest}
  \end{flushleft}
  \vspace*{-1.2cm}
\begin{center}
 \includegraphics[width=\hsize, clip]{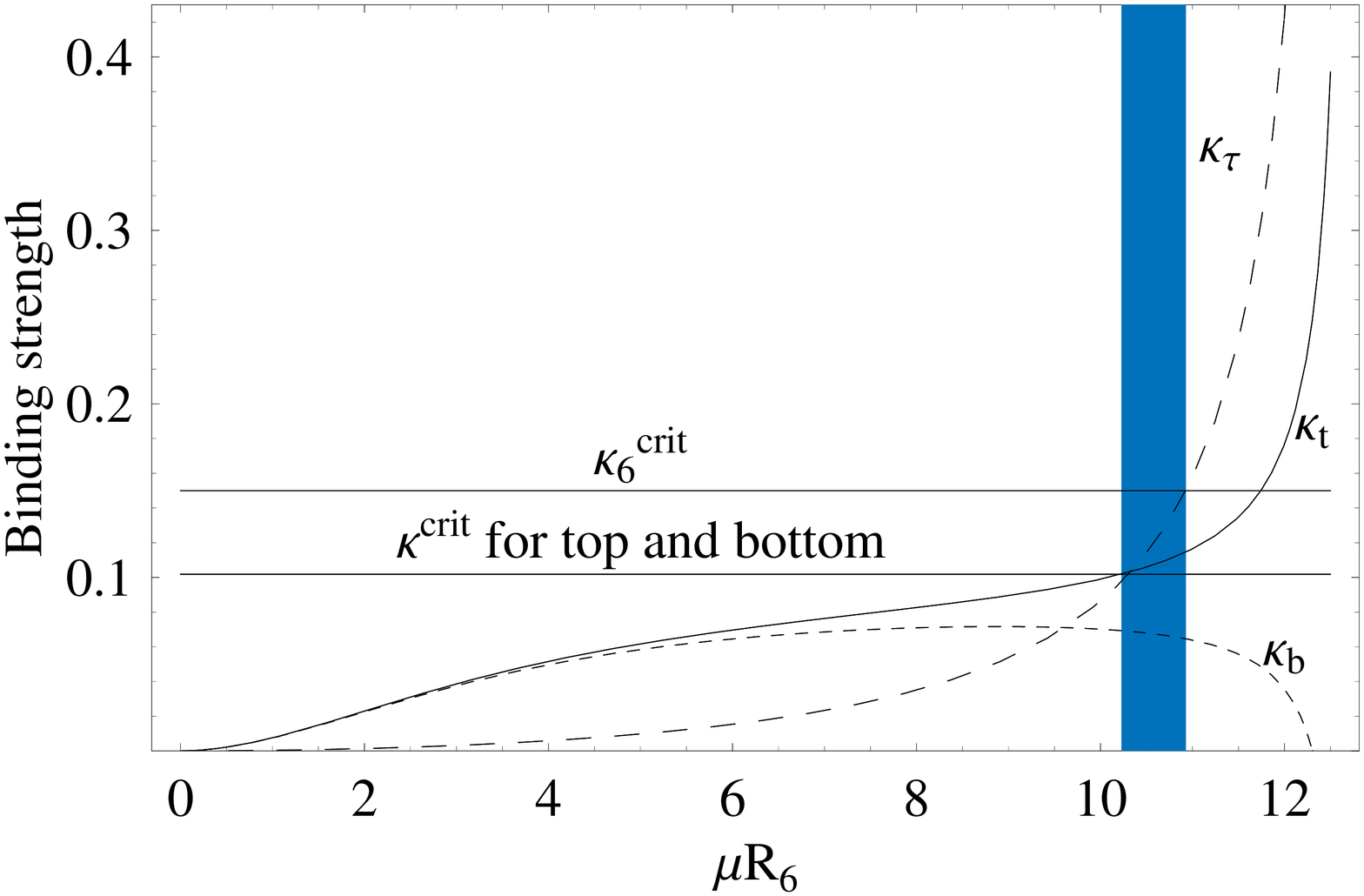}
\end{center}
\end{minipage}
\begin{minipage}{0.5\hsize}
 \begin{flushleft}
    \hspace*{1mm} {(d) \hspace*{6mm} $R_6^{-1}=1\,\TeV$, 4-th}
  \end{flushleft}
  \vspace*{-1.2cm}
\begin{center}
 \includegraphics[width=\hsize, clip]{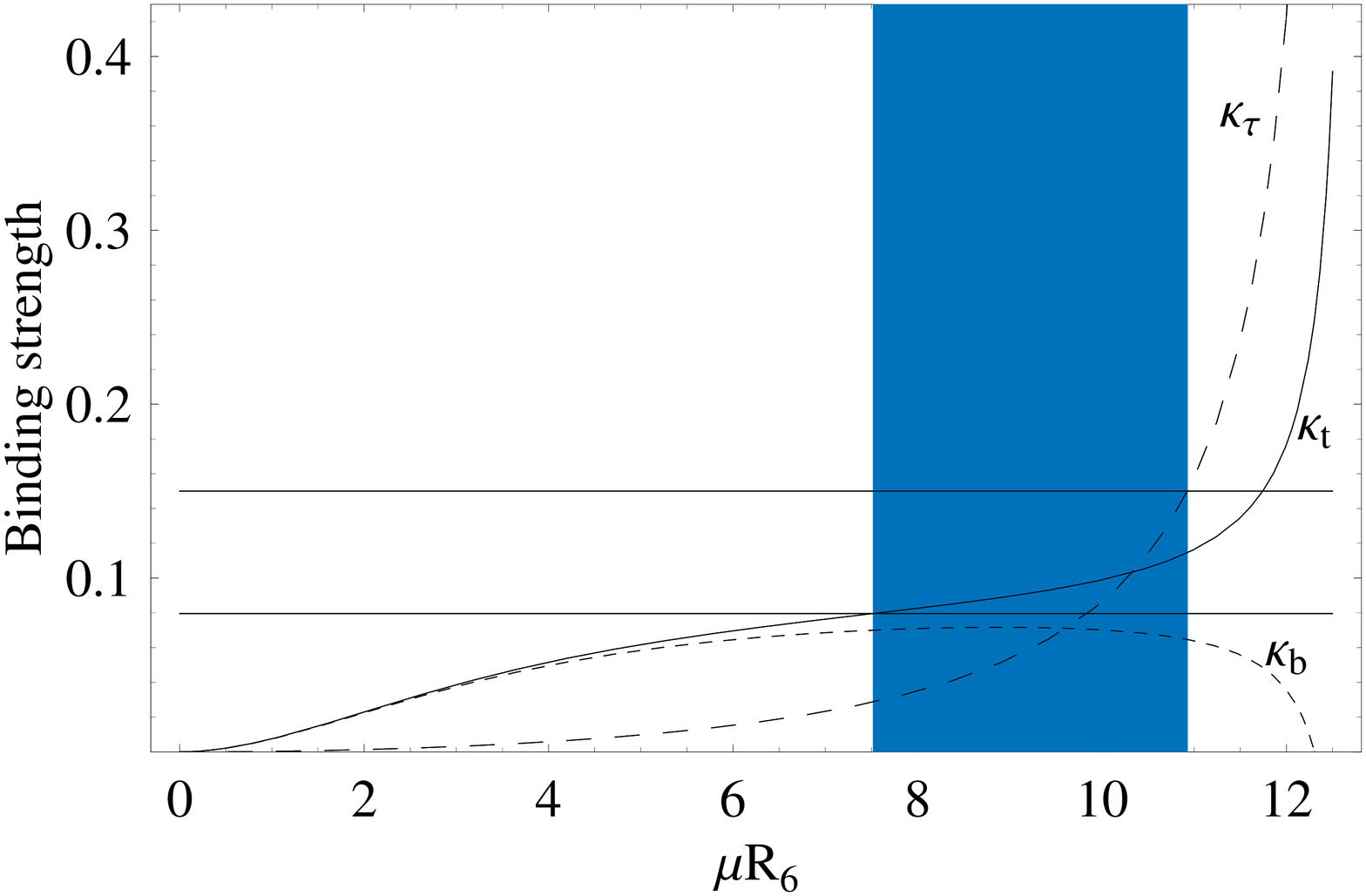}
\end{center}
\end{minipage}
\end{tabular}
\caption{Binding strengths for tau, bottom and top.
          In (a) and (b) shown are the binding strengths $\kappa_\tau,\kappa_t$ and $\kappa_b$ for tau, 
         top and bottom, respectively  on the 5-brane with $R_6^{-1}=10\,\TeV$, while
         in (c) and (d) shown are those with $R_6^{-1}=1\,\TeV$.
         The upper horizontal line in the figures is the critical binding strength for tau, 
         $\kappa_6^{\rm crit}=0.15$
         (nonlocal gauge fixing method).
         The lower line is the one for top and bottom, $\kappa^{\rm crit}$
         derived by using the upper bound of $c$ 
         (the coefficient of the dimensionless four fermion coupling) : (a),(c) are
         those by the lowest KK-mode only, while (b),(d) are those by
         the summation till the 4-th KK-modes.
         The shaded regions which are equivalent to the corresponding shaded regions in Fig.~\ref{crit-MAC}
         are the tMAC scales: 
         $\Lambda_{\rm tM} R_6 = {\rm (a)} 10.2-10.8 \,, {\rm (b)} 7.8-10.8 \,,
         {\rm(c)} 10.3-11 \, {\rm and} \, {\rm(d)} 7.5-11$.
         }
\label{MAC-nonlocal}
\end{figure}

We now demonstrate that the tMAC scale does exist when the gluons in the 8-dimensional bulk give 
the induced four-fermion interactions 
in Eqs.~(\ref{lowest-g}), (\ref{4th-g}) (vertical stripe pattern regions Fig.~\ref{crit-MAC}):
\begin{eqnarray}
 0<g_6^{\rm induced} =g_6^{(8)}<1.22 & \hspace*{2mm} & (\text{by the lowest KK-mode}),  \nonumber \\
 0.18<g_6^{\rm induced} =g_6^{(8)}<1.42 & \hspace*{2mm} & (\text{by the 4-th KK-modes}).
\end{eqnarray}
This implies :
\begin{eqnarray}
  \kappa_t(\Lambda_{\rm tM}) >\kappa^{\rm crit} 
   >\kappa_6(g_6)\Big|_{g_6=1.22} =0.10& \hspace*{2mm} & (\text{the lowest KK-mode}), \nonumber \\
\kappa_t(\Lambda_{\rm tM}) >\kappa^{\rm crit} 
 >\kappa_6(g_6)\Big|_{g_6=1.42}=0.08  & \hspace*{2mm} & (\text{sum to the 4-th KK-modes}).
\label{induced 8}
\end{eqnarray}
 From Fig.~\ref{fig:tMAC0} or Fig.~\ref{MAC-nonlocal}
  we can see that  this is fulfilled for
\begin{equation}
 \Lambda_{\rm tM} >
 \begin{cases}
 10.2 R_6^{-1} &\hspace*{2mm}(\text{lowest only}) \\[3mm]
 7.8 R_6^{-1} &\hspace*{2mm}(\text{sum to the 4-th}), 
 \end{cases}
\end{equation} 
for $R_6^{-1}=10\,\TeV $and 
\begin{equation}
 \Lambda_{\rm tM} >
 \begin{cases}
 10.3 R_6^{-1} &\hspace*{2mm}(\text{lowest only}) \\[3mm]
 7.5 R_6^{-1} &\hspace*{2mm}(\text{sum to the 4-th}),
 \end{cases}
\end{equation} 
for $R_6^{-1}=1\,\TeV$.
Then this time there is an overlap 
with the scale required by the no tau condensation, Eq.(\ref{notaucond}):
\begin{equation}
 \Lambda_{\rm tM} < \Lambda_{\tau} =
 \begin{cases}
 10.8 R_6^{-1}&(\text{for}\,R_6^{-1}=10\,\TeV) \\[3mm]
 11.0 R_6^{-1}&(\text{for}\,R_6^{-1}=1\,\TeV).
 \end{cases}
\end{equation}
Thus the tMAC scale does exist:
\begin{equation}
 \Lambda_{\rm tM} R_6 =
 \begin{cases}
 10.2-10.8 &\hspace*{2mm}(\text{lowest only}) \\[3mm]
 7.8-10.8 &\hspace*{2mm}(\text{sum to the 4-th}), 
 \end{cases}
 \label{tMAC10TeV}
\end{equation} 
for $R_6^{-1}=10\,\TeV$ and 
\begin{equation}
 \Lambda_{\rm tM} R_6 =
 \begin{cases}
 10.3-11.0 &\hspace*{2mm}(\text{lowest only}) \\[3mm]
 7.5-11.0 &\hspace*{2mm}(\text{sum to the 4-th}),
 \end{cases}
 \label{tMAC1TeV}
\end{equation} 
for $R_6^{-1}=1\,\TeV$.

As an illustration  we show in Fig.~\ref{crit-MAC} 
the region of Eq.(\ref{induced 8}) and Eq.(\ref{notautcond}) by
the vertical stripe pattern region and by the 
horizontal stripe pattern region, respectively for $R_6^{-1}= 10 \,{\rm TeV}$ 
(A similar result is obtained for $R_6^{-1}= 1 \,{\rm TeV}$). The tMAC scale defined 
by Eq.(\ref{tMAC1}) is  
the overlap region of these two, which does exist for the case of the induced four-fermion
coupling
$g_6^{(8)}$. 
In Fig.~\ref{MAC-nonlocal} we indicate the tMAC scale, Eq.(\ref{tMAC1}), as the shaded region which is the
overlap region of  Eq.(\ref{induced 8}) and Eq.(\ref{notautcond}) for $R_6^{-1}= 10\, {\rm TeV}$ and
$R_6^{-1}= 1\, {\rm TeV}$. Thus we conclude that tMAC scale does exist.

As to the concrete value of the tMAC scale, there is some ambiguity.
Without knowing further information on the recoil effects of the brane, we may make a best compromise
by taking a conservative
estimate of the sum of the KK modes of the bulk gluons up till the 4th KK modes, which is quite stable
against summing more KK modes contributions in a naive way (see Appendix~\ref{app1}). Then
our conservative estimate of the tMAC scale is
\begin{equation}
 \Lambda_{\rm tM} R_6 =
 \begin{cases}
 7.8-10.8 &\hspace*{2mm}(R_6^{-1}=10\, \TeV) \\[3mm]
 7.5-11.0 &\hspace*{2mm}(R_6^{-1}=1\, \TeV).
 \end{cases}
 \label{conservative}
\end{equation}

Note however that
we have a freedom of tuning the brane position to reduce the induced four-fermion interaction at our
disposal, so that we can always adjust the tMAC scale to the high end of the above estimate:
\begin{equation}
\Lambda_{\rm tM} R_6 =
 \begin{cases}
 10.8 &\hspace*{2mm}(R_6^{-1}=10\,\TeV) \\[3mm]
 11.0 &\hspace*{2mm}(R_6^{-1}=1\,\TeV).
 \end{cases}
 \label{tunedtMAC}
\end{equation}.

\section{Predictions of $m_t$ and $m_H$}
\label{predict mass}

We now calculate masses of the top quark $m_t$ and the Higgs boson $m_H$.
Since it is rather complicated to do the numerical analysis of the TMSM
with extra dimensions using the method of the original MTY~\cite{MTY89}
in 4 dimensions based on the SD equation and the Pagels-Stokar formula, we here follow  
the procedure of ACDH~\cite{ACDH:2000hv} 
(see also \cite{Kobakhidze:1999ce}) and Ref.~\cite{HTY2003} where
the 6-dimensional TMSM was rewritten into the form of the 6-dimensional
SM with the compositeness condition {\' a} la Bardeen-Hill-Lindner (BHL)~\cite{BHL:1989ds}, which was then analyzed in the truncated KK effective theory 
by the renormalization group equations (RGEs) for top-yukawa coupling $y_t$ and 
Higgs quartic coupling $\lambda_H$ 
with the compositeness condition.

In Refs.~\cite{ACDH:2000hv,HTY2003} which have no explicit four-fermion interactions in 6 dimensions, the meaning of the compositeness condition was rather
obscure. In contrast, in our case having explicit four-fermion interaction we 
can formulate straightforwardly the compositeness condition in the 6-dimensional TMSM in precisely the same manner as in the  BHL for the 4-dimensional model. 
Note that the compositeness scale $\Lambda$, which is the scale of the induced four-fermion interactions, namely the  compactification
scale of the seven-th and eighth dimensions  
$\Lambda=R^{\rm -1}\equiv R_7^{\rm -1}=R_8^{\rm -1}$, 
is  not an arbitrary parameter in
contrast to ACDH~\cite{ACDH:2000hv} and Ref.\cite{Kobakhidze:1999ce}
but is identified with the tMAC scale, 
$\Lambda =\Lambda_{\rm tM}$ as in Ref.~\cite{HTY2003}. Then the
compositeness conditions read 
\begin{equation}
 y_t(\mu) \to \infty , 
 \quad \frac{\lambda_H(\mu)}{y_t(\mu)^4} \to 0  \quad
 (\mu \to \Lambda=\Lambda_{\rm tM})\, , \label{comp-cond}
\end{equation}
where $R^{-1}_6 \le \mu \le \Lambda=R^{-1} = \Lambda_{\rm tM}$.

In the truncated KK effective theory the RGEs for the  gauge couplings are 
Eq.(\ref{rge_4d}) and Eq.(\ref{rge-1}) with Eqs.(\ref{bKK-qcd}),(\ref{b2KK}),
(\ref{byKK}).
Similarly, the RGEs for $y_t$ and $\lambda_H$ are given by
\begin{align}
&(4\pi)^2 \mu \frac{d y_t}{d \mu} = \beta_{y_t}^{\rm SM} + \beta_{y_t}^{\rm KK} 
 \label{rge_y} \\[2mm]
&(4\pi)^2 \mu \frac{d \lambda_H}{d \mu} =
  \beta_{\lambda_H}^{\rm SM} + \beta_{\lambda_H}^{\rm KK},  \label{rge_lam}
\end{align}
where,
\begin{eqnarray}
\beta_{y_t}^{\rm SM} &=& 
  y_t \left[\,\left(3+\frac{3}{2}\right)y_t^2 
  - 8 g_3^2 - \frac{9}{4} g_2^2 
  -\frac{17}{12} g_Y^2 \,\right] , \\[2mm]
\beta_{y_t}^{\rm KK} &=& 
 \left( 6 \NKKf +\frac{3}{2} \NKKs \right)\,y_t^3 
 \nonumber \\ && \quad
 -\NKKg \left( 8 g_3^2 + \frac{9}{4} g_2^2 + \frac{17}{12} g_Y^2 \right)\,y_t
  \nonumber \\ && \quad 
 - \delta \NKKb
 \left(\frac{4}{3} g_3^2 - \frac{3}{8} g_2^2 - \frac{1}{72} g_Y^2 \right)\,y_t,
 \label{rge_y-ED} \\
\beta_{\lambda_H}^{\rm SM} &=&
  12 \left(\lambda_H y_t^2-y_t^4\right) + 12\lambda_H^2 \nonumber \\ && \quad 
  +\frac{3}{4}(3 g_2^4 + 2g_2^2 g_Y^2 + g_Y^4) \nonumber \\ && \quad
  -3(3g_2^2+g_Y^2)\lambda_H, \\
\beta_{\lambda_H}^{\rm KK} &=&
   24 \NKKf \left(\lambda_H y_t^2-y_t^4\right) 
 + 12 \NKKs \lambda_H^2  \nonumber \\ && \hspace*{-0.5cm}
 + \NKKg \left[\,
   \frac{3}{4}\left(3 g_2^4 + 2g_2^2 g_Y^2 + g_Y^4\right)
 - 3(3g_2^2+g_Y^2)\lambda_H\,\right] \nonumber \\ && \hspace*{-0.5cm}
 + \frac{\delta}{4} \NKKb (3 g_2^4 + 2g_2^2 g_Y^2 + g_Y^4). 
\end{eqnarray}
By solving these RGEs with inputs Eq.(\ref{inputs}) and the compositeness 
condition Eq.~(\ref{comp-cond}), we determine the running of $y_t (\mu)$ and
$\lambda_H (\mu)$ and predict $m_t$ and $m_H$ by the condition:
\begin{equation}
 m_t =  v \, \frac{y_t(m_t)}{\sqrt{2}}, \quad
 m_H = v \, \sqrt{\lambda_H(m_H)} , \label{rge-analysis}
\end{equation}
where $v=246 \, \GeV$.

We now present our main result: 
From the analysis in Sec.~\ref{gnjl in 6d} we predict $m_t$ and $m_H$
for the conservative estimate of $
\Lambda_{\rm tM}$ in Eq.(\ref{conservative}):
\begin{eqnarray}
 m_t &=& 178 - 187 \, \GeV ,
 \nonumber \\
 m_H &=& 183 - 199 \, \GeV ,
 \label{topmass-10TeV}
\end{eqnarray}
for $\Lambda_{\rm tM} = (7.8-10.8) \, R^{-1}_6 \,(R^{-1}_6 = 10\TeV)$ , and
\begin{eqnarray}
 m_t = 177 - 187 \, \GeV ,
\nonumber \\
 m_H = 186 - 207 \, \GeV ,
 \label{topmass-1TeV}
\end{eqnarray}
for $\Lambda_{\rm tM} = (7.5-11.0)\, R^{-1}_6 \,(R^{-1}_6 = 1\,\TeV)$.

It is remarkable that our top mass prediction 
 \begin{equation}
 m_t = 177 - 187 \, {\rm GeV} \quad (R_6^{-1}=1-10 \, {\rm TeV}) . 
 \label{topmassprediction}
 \end{equation}  
is quite stable against changing 
the compactification scale of the 5-th and 6-th dimensions $R^{-1}_6$. 
\footnote{
Naively, one might think $m_t$ for $R_6^{-1}=1\, {\rm TeV}$ is larger than $m_t$ for $R_6^{-1}=10\, {\rm TeV}$, 
because the compositeness scale for the former case is lower than the latter.
However, the naive guess from the 4-dimensional RGE analysis is not applicable, 
since   
the KK-modes contributions other than
the 4-dimensional SM contributions are operative  in the different energy region
for both cases $R_6^{-1}=1\, {\rm TeV}$ and $R_6^{-1}=10\, {\rm TeV}$.
}
This $m_t$ is consistent with the  new experimental value 
(pole mass)~\cite{CDF/D0:2004}, $178.0 \pm 4.3 \, {\rm GeV}$,\footnote{
After submitting the manuscript, we were informed of the latest experimental results
with somewhat smaller values, $174.3 \pm 3.4 \, {\rm GeV}$~\cite{unknown:2005cg}, 
$172.7 \pm 2.9 {\rm GeV}$~\cite{Group:2005cc},
which are based on the published  Run I
and the preliminary Run II results of the Tevatron. 
}
 and the corresponding
 $\overline{\rm MS}$-mass, $m_t^{\overline{\rm MS} }= 169.8 \pm 4.1\, {\rm GeV}$,   obtained through a formula~\cite{Marciano:1989xd}.

As to the Higgs mass prediction, our conservative estimate implies
 \begin{equation}
 m_H = 183 - 207 \, {\rm GeV} \quad (R_6^{-1}=1-10 \, {\rm TeV}) . 
 \end{equation}  
 This Higgs boson mass prediction, 
somewhat similar to that of Ref. \cite{HTY2003}, $m_H=176-188$ GeV, 
is characteristically smaller 
than that of the typical dynamical EWSB models like technicolor.
On the other hand, the value is substantially 
larger than that of 
typical supersymmetric models, $m_H \lessim 130$ GeV (MSSM) or 
$m_H \lessim 150$ GeV (NMSSM).
Thus the present scenario is clearly distinguished from 
many of the typical models beyond the SM, either dynamical or SUSY models,
simply through the Higgs mass observation.
The Higgs boson of this mass range
decays into weak boson pair almost 100\%.
It will be immediately discovered 
in $H \to WW^{(*)}/ZZ^{(*)}$ once the LHC starts.
\\

{\it Some comments are in order}:

As we discussed in Sec.\ref{4fermi in 6d}, there is a possibility that
the recoil effects of the brane reduce higher KK modes drastically, in which 
case only the lowest KK-mode contribution, instead of sum up till the 
4th KK modes,  may be the relevant contribution to the induced four-fermion coupling $g_6^{\rm induced}$.
If we take the $\Lambda_{\rm tMAC}$ values for only the lowest KK-mode as given
in Eqs.(\ref{tMAC10TeV}) and (\ref{tMAC1TeV}) 
instead of the conservative estimate of
$\Lambda_{\rm tM}$ in Eq.(\ref{conservative}), the prediction is:
\begin{equation}
 m_t = 178 - 180 \, \GeV, \hspace*{4mm} m_H = 183 - 186 \, \GeV 
 \label{lowesttop10TeV}
\end{equation}
for $\Lambda_{\rm tM} = 10.2-10.8\,R^{-1}_6\,(R^{-1}_6 = 10\, \TeV)$ ,
\begin{equation}
 m_t = 177 - 179 \, \GeV, \hspace*{4mm} m_H = 186 - 190 \, \GeV
 \label{lowesttop1TeV}
\end{equation}
for $\Lambda_{\rm tM}= 10.3-11.0\,R^{-1}_6(R^{-1}_6 = 1\, \TeV)$.
The prediction becomes somewhat more restricted for the mass range.

If we further exploited the freedom of the brane position as to tune the 
induced four-fermion coupling as given by Eq. (\ref{tunedtMAC}), then we may
pinpoint the prediction to the lower end values in Eqs.(\ref{topmass-10TeV}),(\ref{topmass-1TeV}),(\ref{lowesttop10TeV}) and 
(\ref{lowesttop1TeV}):
 \begin{equation}
 m_t = 177-178 \, \GeV \quad (R^{-1}_6 = 1-10\, \, \TeV) ,
 \end{equation}
 \begin{equation} 
 m_H = 183-186 \, \GeV  \quad (R^{-1}_6 = 1-10\, \,\TeV) .
 \label{lowerendhiggs}
 \end{equation}
Note that the above lower end values of the prediction are not altered even if
we included  20\% errors of the possible ambiguity of the SD equation 
we mentioned 
earlier. Considering the top mass prediction should be close to the reality, the most
plausible value of the Higgs mass prediction in our model would be such lower end values Eq.(\ref{lowerendhiggs}).

For comparison, we may present values calculated when our analysis is performed in 
the Landau gauge fixing as in Ref.~\cite{HTY2003}, although the Landau gauge 
analysis is less reliable than that in the nonlocal gauge as we discussed before.
The result actually is not changed so much:
The tMAC scale is 
\begin{equation}
 \Lambda_{\rm tM} R_6 =
 \begin{cases}
 8.3-10.5 &\hspace*{2mm}(\text{lowest only}) \\[3mm]
 6.5-10.5 &\hspace*{2mm}(\text{sum till the 4-th}), 
 \end{cases}
\end{equation} 
for $R_6^{-1}=10\, \TeV$ and 
\begin{equation}
 \Lambda_{\rm tM} R_6 =
 \begin{cases}
 8.0-10.6 &\hspace*{2mm}(\text{lowest only}) \\[3mm]
 6.5-10.6 &\hspace*{2mm}(\text{sum till the 4-th}),
 \end{cases}
\end{equation} 
for $R_6^{-1}=1\, \TeV$. Note that the lower end value for the sum till the
4-th KK 
modes is the same for   $R_6^{-1}=10\, \TeV$ and $R_6^{-1}=1\, 
\TeV$, which is determined by the requirement of no bottom condensation since in this case the $\kappa^{\rm crit}$ is lower than that in 
the nonlocal gauge (the value given in Fig.\ref{MAC-nonlocal}).
Accordingly, the masses for top and Higgs are predicted as 
\begin{equation}
 m_t = 179 - 192 \, \GeV, \hspace*{4mm} m_H = 187 - 211 \, \GeV 
\end{equation}
for $\Lambda_{\rm tM} = 6.5-10.8\,R^{-1}_6\, (R^{-1}_6 = 10\, \TeV)$, and
\begin{equation}
 m_t = 178 - 192 \, \GeV, \hspace*{4mm} m_H = 187 - 218 \, \GeV
\end{equation}
for $\Lambda_{\rm tM}= 6.5-11.0\,R^{-1}_6 \,(R^{-1}_6 = 1\, \TeV)$,
which are compared with Eq.(\ref{topmass-10TeV}) and (\ref{topmass-1TeV}), respectively.

\section{Summary}
\label{summary}

We have proposed a version of the Top Mode Standard Model (TMSM) 
in six dimensions
(5-brane in the eight-dimensional bulk),
with the third generation quarks/leptons and the $SU(2)_L \times U(1)$ gauge 
bosons living on the 5-brane with the 5-th and 6-th dimensions compactified on 
$T^2/Z_2$ with TeV scale, $R_5^{-1}=R_6^{-1}=1-10 \, \TeV$, 
while the $SU(3)$ gluons, 
 living in the eight-dimensional bulk with
the 6-th and 7-th dimensions compactified on $T^2/Z_2$ with 
yet higher scale, $R_7^{-1}=
R_8^{-1}=\Lambda \gg R_6^{-1}$,  give
rise to induced four-fermion interactions of top and bottom (but not of tau)
on the 5-brane. The first and second generations are living in four dimensions. Having such a four-fermion 
interactions induced by the bulk gluon KK modes in addition to the  Standard Model gauge interactions on the 5-brane,  the model for top/bottom 
takes the form of the
6-dimensional gauged NJL model whose critical line is given by
Eq.(\ref{criticalline1}) with $D=6$. 
We have shown that such an induced four-fermion coupling is well above
the critical line $g_6^{\rm induced}=g_6^{(8)} >1/2$, and in fact 
strong enough as to 
trigger the top condensate without bottom and tau condensates. Namely, there exists an energy region
$\mu$ (tMAC scale) satisfying the condition Eq.(\ref{tMAC1}), see the shaded region in Figs. \ref{crit-MAC} and \ref{MAC-nonlocal}.

Here we note that our estimation of the induced four-fermion interactions crucially depends on the
existence of UVFP~\cite{HTY:2000uk,Agashe:2000nk,Kazakov:2002jd,Dienes:2002bg}.
Although existence of such a UVFP is still in controversy, pro and con,
in lattice studies~\cite{kawai} and other nonperturbative methods~\cite{Gies:2003ic}, its existence will result in resolving 
a possible conflict with arguments of the perturbative
unitarity~\cite{Chivukula:2003kq} which presume no such a UVFP.
Moreover, the brane 
fluctuation strongly suppresses higher KK modes as in Eq.(\ref{gauge-sup})~\cite{Bando:1999}, which makes the ``divergence'' of the summation of KK modes merely
superficial. This is another source to avoid conflict with the 
perturbative unitarity arguments. 
We also note that as was explicitly checked in Ref.\cite{HTY2003} the  
KK modes summation is fairly independent of the truncation scheme for
$D=6$ and $D=8$, though not for $D=10$.

In the truncated KK effective theory~\cite{DDG} we employed in this paper,
the $SU(3)\times SU(2)\times U(1)_Y$ SM gauge couplings on the 5-brane are
``strong'' enough to trigger the top quark condensate but still ``weak'' 
enough not to destroy the perturbative picture completely: The binding 
strength is given by $\kappa_i = \cal{O}$$ (0.1)$ $(i=t,b,\tau)$ for the relevant energy region $\mu R_6 < 10$ (see Fig.~\ref{MAC-nonlocal}), which are much smaller than the naive dimensional analysis
$\kappa =\cal{O}$$(1)$. Thus the gauge theory (including $U(1)$) on the 5-brane   
also is not obviously in conflict with the perturbative unitarity. 

It should be emphasized that our compactification of the 8-dimensional bulk
into the 5-brane, $D=8 \rightarrow D=6$, is on $T^2/Z_2$ instead of $T^2$, 
which leaves us with  parameters, the brane position $x_{70},\, x_{80}$,
to tune the four-fermion coupling close to the critical line so that
 the dynamical mass of the top quark,
 which is otherwise on the order of cutoff $\Lambda$,
 can be kept on the weak scale order in the SD gap equation:
 $m_t \sim v = 246 \GeV\ll \Lambda$. Such a freedom corresponds to tuning
 the VEV of the composite Higgs,
$v \ll \Lambda$, in the BHL formulation based on the
RGE's plus compositeness conditions.

We then calculated based on the BHL formulation the predicted values:
\begin{equation}
m_t = 178 - 187 \,\GeV, \quad  m_H = 183 - 207 \,\GeV.
\end{equation}
 The top mass prediction is
consistent with the experimental value (see the discussions below Eq. (\ref{topmassprediction})). The Higgs boson mass prediction is a rather 
characteristically small value compared with those in other strongly coupled
Higgs models like technicolor which are usually
larger than that.
On the other hand, the value is substantially 
larger than that of 
typical supersymmetric models, $m_H \lessim 130$ GeV (MSSM) or $m_H \lessim 150$ GeV (NMSSM).
Thus the present scenario is clearly distinguished from 
many of the typical models beyond the SM
simply through the Higgs mass observation.
The Higgs boson of this mass range
decays into weak boson pair almost 100\%
and  will be immediately discovered 
in $H \to WW^{(*)}/ZZ^{(*)}$ once the LHC starts.

Several comments are in order: 
\begin{itemize}
 \item In this paper we discussed mass of the top quark as the origin of the masses of W and Z bosons
and the composite Higgs. What about the mass of other quarks and leptons?
\begin{itemize}
\item Bottom mass\\
In the original TMSM~\cite{MTY89}, the bottom mass must come 
from $G_{ts}$-term in Eq. (\ref{eq:tb-4fermi}) ($G^{(2)}$ term in Ref.\cite{MTY89})
:
\begin{equation}
 G_{tb}\left(\epsilon^{ik}\epsilon_{jl}\bar\psi_L^i {\psi_R}_j
     \bar\psi_L^k {\psi_R}_l\right)
   + h.c.,
\end{equation} 
which explicitly breaks the Peccei-Quinn symmetry and yields  $m_b = - G_{tb} \VEV{\bar t t}$ after
the top condensation
takes place. Were it not for the $G_{ts}$ term, the bottom condensate due to strong
$G_b$ term in Eq.(\ref{eq:tb-4fermi}) would lead to
the visible axion (2nd paper in Ref.\cite{MTY89}) which is already ruled out. 
As was emphasized in Ref.\cite{Tanabashi:1989sz} 
the $G_{ts}$ term does not arise from the massive vector meson 
exchange model and hence from the gauge interaction. However it was pointed out~\cite{Yamawaki:1990ue}
that the instanton effects can give rise to such a term, although it turned out very small in the original 
TMSM with large cutoff~\cite{Tanabashi:1992rn}. 
It was argued~\cite{Hill:1994hp,He:2001fz}, however,  in the topcolor scenario with much smaller cutoff 
scale compared with the original TMSM, that such instanton effects 
can produce a reasonable amount of mass for the bottom.

In the case at hand, we may naively guess from the $D=4$ color instanton that the bulk 
gluon instanton living in $D=8$ bulk would give the  $D=6$ $G_{ts}$-like four-fermion interaction
on the 5-brane, although little is known 
about the higher dimensional instantons (see e.g., Ref.~\cite{Hill:2000rr}).

\item Tau mass \\
In the original TMSM~\cite{MTY89}, the tau mass also comes from a term similar to
the $G_{ts}$ term with the $\psi=(t,b)$ replaced by $\psi=(\nu_\tau, \tau^-)$ for the
one pair of $\bar \psi_L \psi_R$. In the present model, however, without introducing 
ad hoc four-fermion interactions, 
we would need some
larger picture such as the Pati-Salam gauge unification between bottom and tau.

\item 1st and 2nd generation masses\\
There are various possible ways to communicate the top condensate with the mass operator of
the 1st and 2nd
generations: The simplest one would be the one similar to the extended technicolor
 (ETC)\cite{Dimopoulos:1979es}
 with the role
of the technifermion now replaced by the top quark, namely through the horizontal gauge
interaction on the 3-brane. Since the anomalous dimension of the top quark condensate of our model 
in terms of the 4-dimensional language is close to 2, 
$\gamma_m \simeq 2$~\cite{MTY89,Yamawaki96}, much larger than even the walking 
technicolor ($\gamma_m \simeq 1$)~\cite{Yamawaki:1985zg},
there is no such conflict to the Flavor-Changing Neutral Currents (FCNC) 
as that in the ETC.
\end{itemize}

\item Constraints from the precision experiments\\
The new particles other than the SM particles contributing to the $S$ parameter
are KK modes of the top/bottom which , however, 
are vector-like and hence yield little contributions to the $S$ parameter~\cite{Maekawa:1994yd}.
As to the $T$ parameter or $\Delta \rho$,
the summation of KK modes below the cutoff $\Lambda$ contributes to
$\Delta\rho$ as $\Delta \rho \sim 10 (M_W R)^2$~\cite{Nath:1999fs} which 
would unfavor the lower $R^{-1}\sim 1 {\rm TeV}$ similarly to  
the previous model with $D=8$~\cite{HTY2003}.

\item UV sensitivity\\
Our model is based on the dynamics of 6-dimensional gauged NJL model, but when $N_c=3, N_f=2$, 
       the 6-dimensional gauged NJL model is non-renormalizable even in the nonperturbative sense
discussed in Ref.~\cite{Gusynin:2004jp}. We would need a better-controlled theory beyond $\Lambda$.

\item In this paper, the freedom of the position of the 5-brane in the higher dimensional bulk
      played a central role for consistency in our model. The origin of this degree of freedom
      remains to be investigated in the brane dynamics.
\end{itemize}

\section*{Acknowledgments}
We thank Michio Hashimoto and Masaharu Tanabashi
for very helpful comments and discussions.
Thanks are also due to Kazuhiko Fujiyama, Masafumi Kurachi and Shinya Matsuzaki
for valuable discussions. 
The work was supported in part by the JSPS Grant-in-Aid for the
Scientific Research (B)(2) 14340072.

\appendix

\section{The possibility of the fermion condensation on the 3-brane}
\label{app1}
We here consider whether or not the fermion fixed on the 3-brane can condense 
by the  four-fermion interactions induced by the bulk gluons in $D (<4)$ dimensions. 
We consider the case with $(D-4)$-compactification  on 
$T^{\delta}/Z_2^k (\delta \equiv D-4 ,k=1,2,\cdots)$ of the  extra dimensions,
with the compactification radii $R_5=R_6=R$,
in which the only gluon propagates.
If the KK-modes effects of the bulk gluons  give rise to four-fermion interactions on the 3-brane, 
such four-fermion interactions may take the form: 
\begin{equation}
\begin{split}
\mathcal{L}_{\rm 4F}&=
\frac{1}{2} \cdot \frac{2^k}{(3\pi R)^{\delta}} \cdot \frac{g_3^2(M_1)}{2M_1^2} \cdot c_4^{(D)} \times (\bar{\psi}\psi)^2\\
&=\frac{G_4^{(D)}}{2N_c}(\bar{\psi}\psi)^2
\end{split}
\end{equation}
where $M_1$ is the mass of the lowest KK-mode $M_1 = R^{-1}$ and $g_3$ is the 
QCD coupling on the 3-brane, and
$c_4^{(D)}$ is the dimensionless coefficients to be estimated below.
The dimensionless four-fermion coupling on the 3-brane is thus
\begin{equation}
\begin{split}
 g_4^{(D)} &= 2^2G_4^{(D)}M_1^2\NDA \\
   &= 2^2 \cdot \frac{2N_c}{2} \frac{2^k}{(3\pi R)^{\delta}} \frac{g_3^2(M_1)}{2M_1^2} c_4^{(D)} \cdot M_1^2\NDA \\
   &= c_4^{(D)}\cdot 2^{(4/2-1)} N_c \cdot  g_3(M_1)^2\NDA.
\end{split} \label{ge-4fermi}
\end{equation}
Let us estimate of  $c_4^{(D)}$  for $D=6,8$ cases.\\

\underline{Case.1 : $D=6$ cases}

First, we consider the $T^2/Z_2$-compactification.
Imposing  periodic boundary condition:
\begin{equation}
\begin{split}
G_\mu(x,x_5,x_6) &= G_\mu(x,x_5+2\pi R,x_6) \\
                 &= G_\mu(x,x_5,x_6+2\pi R),
\end{split} \label{pbc-6}
\end{equation}
and a $Z_2$ condition:
\begin{equation}
G_\mu(x,x_5,x_6) = G_\mu(x,-x_5,-x_6) ,
\end{equation}
we decompose $G_\mu$ as
\begin{equation}
 G_{\mu}(x,x_5,x_6) = \frac{1}{2\pi R} \Bigl[ G_{\mu, 00}(x) + \sqrt{2} \sum_{[n]_1}^{N_{\rm KK}} \tilde{G}_{\mu}^{[1]} 
                  + 2 \sum_{[n]_2}^{N_{\rm KK}} \tilde{G}_{\mu}^{[2]} \Bigl],
\label{6-4-gluon}                  
\end{equation}
where 
\begin{eqnarray}
\tilde{G}_{\mu}^{[1]} &=& G_{\mu, c0}^{[n]_1} \cos\frac{n_1x_5}{R} 
                      + G_{\mu, 0c}^{[n]_1} \cos\frac{n_1x_6}{R} \\
\tilde{G}_{\mu}^{[2]} &=& G_{\mu, cc}^{[n]_2} \cos\frac{n_1x_5}{R}\cos\frac{n_2x_6}{R}                        
                       + G_{\mu, ss}^{[n]_2} \sin\frac{n_1x_5}{R}\sin\frac{n_2x_6}{R}.                                             
\end{eqnarray}
We are interested in the upper bound for $g_4^{(6)}$, which is realized at $x_5=x_6=0$
where these KK-mode induce a four-fermion interaction on the 3-brane as:
\begin{eqnarray}
 \mathcal{L}_{\rm 4F} &=& \frac{1}{2}\frac{2}{(2\pi R)^2}\Biggl[ \sum_{[n]_1}\frac{g_{6,3}^2(M_{[n]_1})}{2M_{[n]_1}^2} \times 2 \nonumber \\
                       && \hspace*{2cm} +2\sum_{[n]_2}\frac{g_{6,3}^2(M_{[n]_2})}{2M_{[n]_2}^2} \times D_{[n]_2} \Biggr] (\bar{\psi}\psi)^2\nonumber \\
                       &=& \frac{1}{2}\frac{2}{(2\pi R)^2}\Biggl[ \sum_{[n]_1}\frac{g_{6,3}^2(M_1)}{2M_1^2}\cdot \Bigl(\frac{M_1}{M_{[n]_1}}\Bigr)^4 \times 2 
                          \nonumber \\
                       &&\hspace*{2cm} + 2\sum_{[n]_2} \frac{g_{6,3}^2(M_1)}{2M_1^2}\cdot \Bigl(\frac{M_1}{M_{[n]_2}}\Bigr)^4 \times D_{[n_2]}\Biggr] (\bar{\psi}\psi)^2 
                           \nonumber \\
                       &=& \frac{1}{2}\frac{g_3^2(M_1)}{2 M_1^2} \times c_4^{(6,k=1)} \times (\bar{\psi}\psi)^2 ,
\end{eqnarray}
where $D_{[n]_i}$ is the degeneracy  having the same $N$ for each $[n]_i$,
\begin{equation}
  [n]_1\equiv [n_1], \quad
  [n]_2\equiv [n_1, n_2], \cdots,
\end{equation}
and $M_{[n]_1}\,,\,M_{[n]_2}$ are
\begin{equation}
 M_{[n]_1}^2 = n^2 M_1^2, \quad M_{[n]_2}^2 = (n_1^2 + n_2^2)M_1^2,\cdots.
\end{equation}
We used the fact that the dimensionless bulk QCD coupling $\hat{g}_{6,3}$ is nearly 
on the UVFP, i.e.,
\begin{equation}
 \hat{g}_{6,3}^2(M_{[n]_i}) = g_{6,3}^2(M_1)\Bigl(\frac{M_1}{M_{[n]_1}}\Bigr)^2,
\end{equation}
and 
\begin{equation}
 g_{6,3}^2(M_1) = \frac{(2\pi R)^2}{2}g_3^2(M_1).
\end{equation}
 From the above we read the coefficient $c_4^{(6,k=1)}$ as
\begin{equation}
 c_4^{(6,k=1)} = 2\sum_{[n]_1} \Bigl(\frac{M_1}{M_{[n]_1}}\Bigr)^4 
               + 2\sum_{[n]_2} D_{[n]_2} \Bigl(\frac{M_1}{M_{[n]_2}}\Bigr)^4,
\end{equation}
which yields the dimensionless four-fermion-coupling $g_4^{(6,k=1)}$ (Eq.~(\ref{ge-4fermi})):
\begin{equation}
 g_4^{(6,k=1)}=c_4^{(6,k=1)}\cdot 2^2 N_c \cdot  g_3(M_1)^2\NDA.
\end{equation}
Note that 
the summation in Eq.~(\ref{6-4-gluon}) stands for the summation of KK-mode whose $\text{mass}^2$ is
\begin{equation}
 M_{[n]_i}^2=\sum_k^i\frac{n_k^2}{R^2}=\frac{N}{R^2} \le \frac{N_{\rm KK}}{R^2}.
\end{equation}

\begin{figure}
\begin{tabular}{cc}
\begin{minipage}{0.5\hsize}
\begin{flushleft}
  \hspace*{0.1cm} {(a) \hspace*{6mm} $\delta=2$ case}
\end{flushleft}
  \vspace*{-0.8cm}
 \includegraphics[width=\hsize,clip]{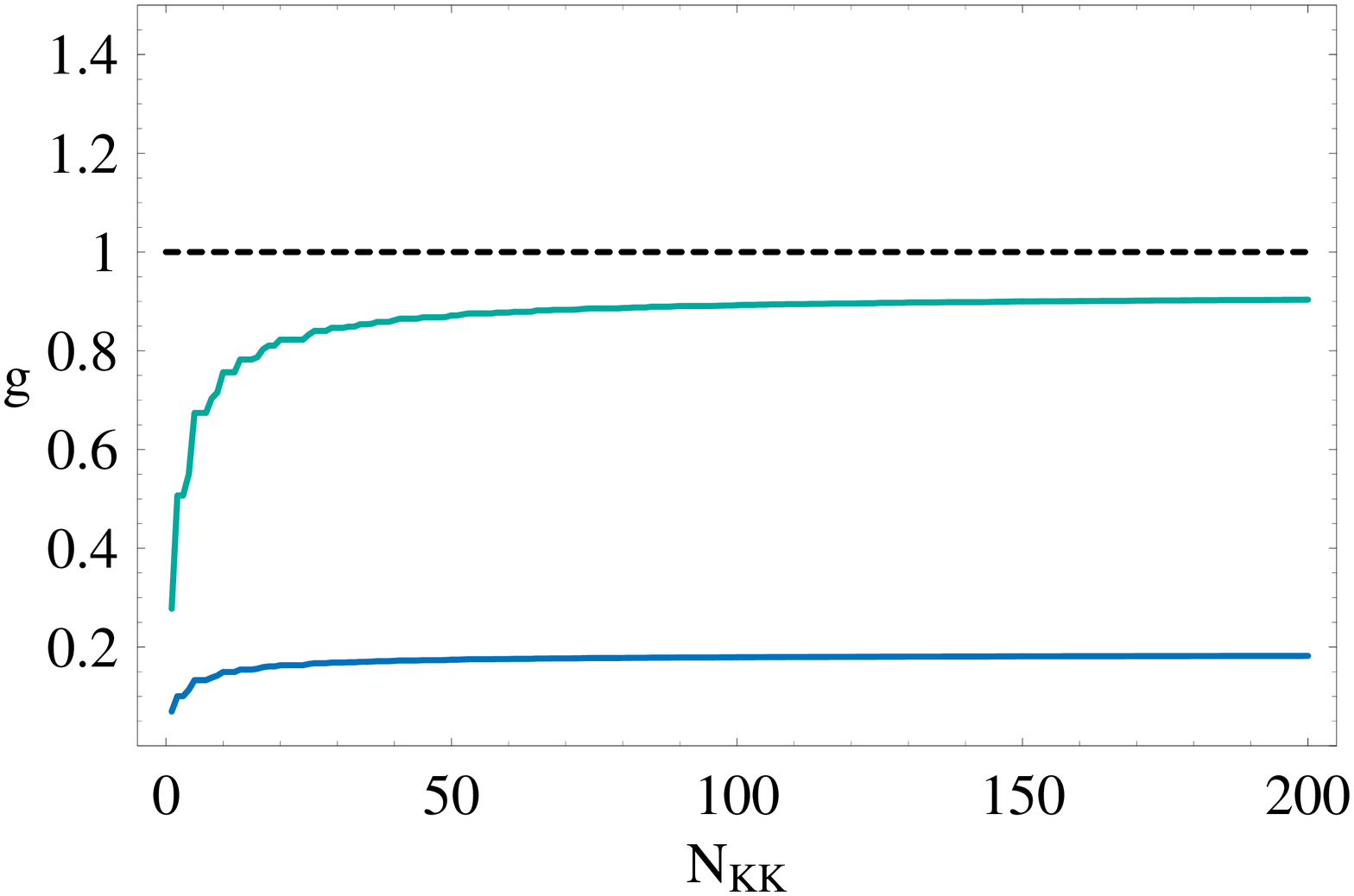}
\end{minipage}
\begin{minipage}{0.5\hsize}
\begin{flushleft}
  \hspace*{0.2cm} {(b) \hspace*{6mm} $\delta=4$ case}
\end{flushleft}
  \vspace*{-0.8cm}
 \includegraphics[width=\hsize,clip]{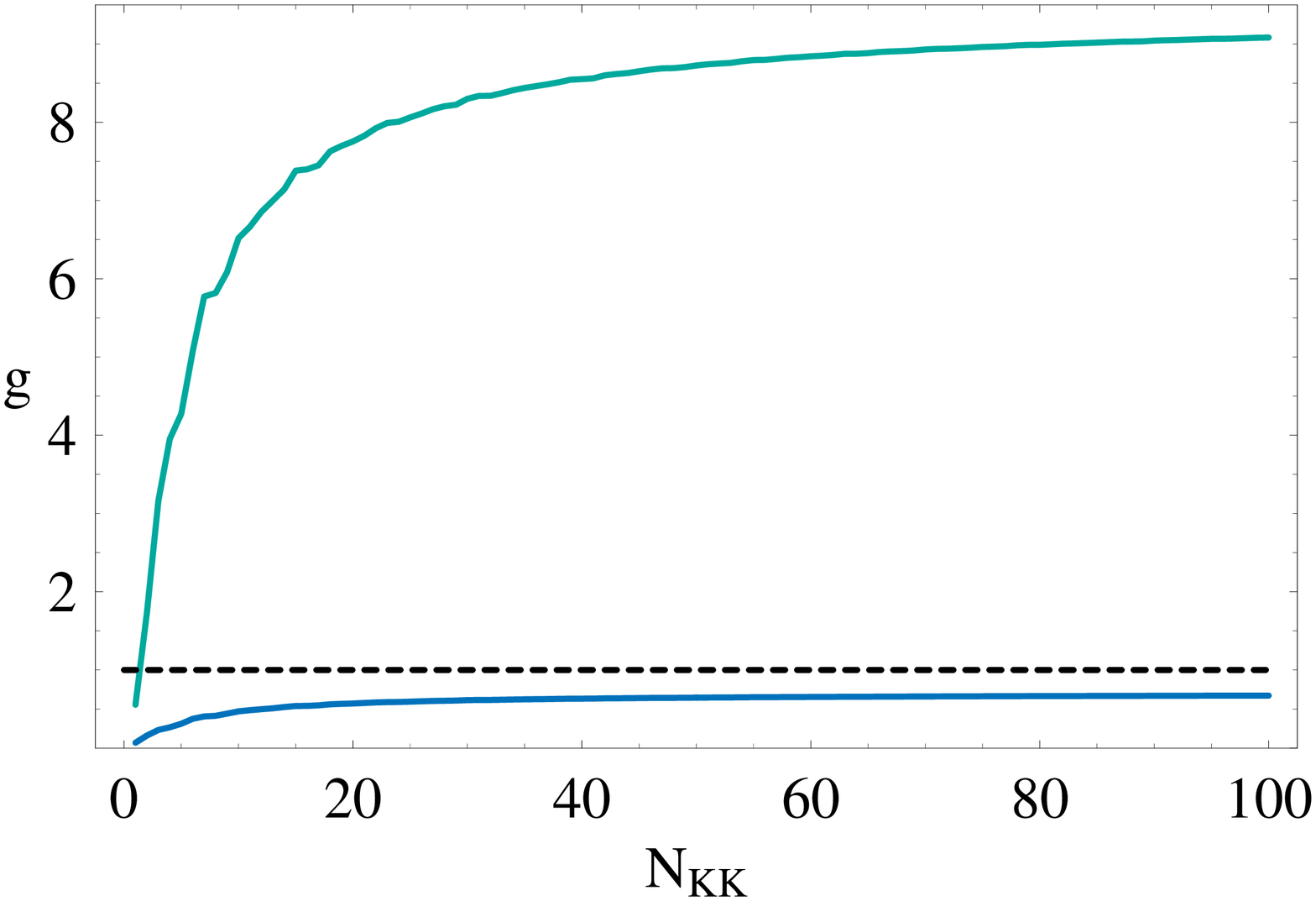}
\end{minipage}
\end{tabular}
\caption{The dimensionless four fermion coupling: $g_4^{(6)}$ on the 3-brane summing up till $N_{\rm KK}$;
         The horizontal dashed line is $g_4^{\rm crit}=1$ on the 3-brane.
         Fig. (a) is for $D=6 (\delta=2)$, and (b) for $D=8 (\delta=4)$.
         The upper lines are $T^{\delta}$-compactification cases with $k=0$ and
         the lower lines are $T^{\delta}/Z_2^k$-compactification cases with $k=1$(for $\delta=2$) and $k=2$(for $\delta=4$).
         }
\label{3brane-4fermi-1}
\end{figure}

Next, we consider $T^2$-compactification ($k=0$), i.e. we impose the periodic boundary condition Eq.~(\ref{pbc-6}) only. 
In this case, the result is independent of the brane position so that we take $x_5=x_6=0$.
In this case 
we must consider $n_1 = \pm1, \pm2, \cdots,$ etc , for $c_{4}^{(6,k=0)}$  and
$c_{4}^{(6,k=0)}$ is given by 
\begin{equation}
 c_{4}^{(6,k=0)} = 2^2\sum_{[n]_1} \Bigl(\frac{M_1}{M_{[n]_1}}\Bigr)^4 
               + 2^3\sum_{[n]_2} D_{[n]_2} \Bigl(\frac{M_1}{M_{[n]_2}}\Bigr)^4 ,
\end{equation}
which yields  
\begin{equation}
 g_4^{(6,k=0)}=c_4^{(6,k=0)}\cdot 2^2 N_c \cdot  g_3(M_1)^2\NDA.
\end{equation}
The numerical estimate of $g_4^{(6,k=0)}$ with $c_4^{(6,k=1)},\,c_{4}^{(6,k=0)}$ in \underline{Case.1} is shown Fig.~\ref{3brane-4fermi-1}(a).
We calculated $g_4^{(6)}$ till $N_{\rm KK}=200$ concretely in this figure.
From this figure, we conclude that the bulk gluons do not give rise to the S$\chi$SB-phase for
the fermions on the 3-brane.\\

\underline{Case.2 : $D=8$ cases}

As in Case.1,  the imposed boundary condition for $T^4/Z_2^2$-compactification case
is also a periodic boundary condition:
\begin{equation}
\begin{split}
G_\mu(x,y,z) &= G_\mu(x,y+2\pi R,z) \\
             &= G_\mu(x,y,z+2\pi R),
\end{split} \label{pbc-8}
\end{equation}
and the $Z_2^2$ condition:
\begin{eqnarray}
G_\mu(x,y,z) &=& G_\mu(x,-y,z) \nonumber\\
             &=& G_\mu(x,y,-z),
\end{eqnarray}
where $y=x_{5,6}$ and $z=x_{7,8}$.
Making a short-hand notation of the $G_{\mu, c000}^{[n]_1}\cos(n_1x_5/R)$ as $G_{\mu, c000}$, etc. 
we write the KK-decomposition of $G_{\mu}(x,y,z)$ as:
\begin{equation}
 G_{\mu}(x,y,z) = \frac{1}{(2\pi R)^2} \Bigl[ G_{\mu, 0000}(x) + \sqrt{2} \sum_{[n]_1}^{N_{\rm KK}} \tilde{G}_{\mu}^{[1]}
                  + 2 \sum_{[n]_2}^{N_{\rm KK}} \tilde{G}_{\mu}^{[2]}
                  + 2\sqrt{2} \sum_{[n]_3}^{N_{\rm KK}} \tilde{G}_{\mu}^{[3]} 
                  + 4 \sum_{[n]_4}^{N_{\rm KK}} \tilde{G}_{\mu}^{[4]} \Bigl],
\label{8-4-gluon}                  
\end{equation}
where
\begin{eqnarray}
 \tilde{G}_{\mu}^{[1]} &=& G_{\mu, c000}^{[n]_1} + G_{\mu, s000}^{[n]_1}
                           + G_{\mu, 0c00}^{[n]_1} + G_{\mu, 0s00}^{[n]_1} \nonumber \\
                       && \hspace*{4mm} + G_{\mu, 00c0}^{[n]_1} + G_{\mu, 00s0}^{[n]_1}
                           + G_{\mu, 000c}^{[n]_1} + G_{\mu, 000s}^{[n]_1}, \\           
 \tilde{G}_{\mu}^{[2]} &=& G_{\mu, cc00}^{[n]_2} + G_{\mu, c0c0}^{[n]_2} 
                           + G_{\mu, c00c}^{[n]_2} + G_{\mu, 0cc0}^{[n]_2} \nonumber \\
                       && \hspace*{4mm} + G_{\mu, 0c0c}^{[n]_2} + G_{\mu, 00cc}^{[n]_2} 
                           + G_{\mu, ss00}^{[n]_2} + G_{\mu, 00ss}^{[n]_2} , \\
 \tilde{G}_{\mu}^{[3]} &=& G_{\mu, ccc0}^{[n]_3} + G_{\mu, cc0c}^{[n]_3} 
                           + G_{\mu, c0cc}^{[n]_3} + G_{\mu, 0ccc}^{[n]_3} \nonumber \\
                       && \hspace*{4mm} + G_{\mu, ssc0}^{[n]_3} + G_{\mu, ss0c}^{[n]_3}
                           + G_{\mu, c0ss}^{[n]_3} + G_{\mu, 0css}^{[n]_3} \\
 \tilde{G}_{\mu}^{[4]} &=& G_{\mu, cccc}^{[n]_4} + G_{\mu, ccss}^{[n]_4} 
                           + G_{\mu, sscc}^{[n]_4} + G_{\mu, ssss}^{[n]_4} . 
\end{eqnarray}
Our interest is an estimation of the upper bound for four-fermion coupling, 
which is realized at the brane positions $x_{50}=x_{60}=x_{70}=x_{80}=0$
where we rewrite these gluons KK-modes effects into the four-fermion interactions on the 3-brane:
\begin{eqnarray}
 \mathcal{L}_{\rm 4F} &=& \frac{1}{2} \frac{2^2}{(2\pi R)^4} 
                          \Biggl[ \sum_{[n]_1}\frac{g_{8,3}^2(M_{[n]_1})}{2M_{[n]_1}^2}\times4 
                          + 2\sum_{[n]_2}\frac{g_{8,3}^2(M_{[n]_2})}{2M_{[n]_2}^2}\times 6 \times D_{[n]_2} \nonumber \\
                       && \hspace*{2cm} + 2^2\sum_{[n]_3}\frac{g_{8,3}^2(M_{[n]_3})}{2M_{[n]_3}^2}\times 4 \times D_{[n]_3} 
                       + 2^3\sum_{[n]_4}\frac{g_{8,3}^2(M_{[n]_4})}{2M_{[n]_4}^2}\times D_{[n]_4} \Biggr] (\bar{\psi}\psi)^2 
                         \nonumber \\
                       &=& \frac{1}{2} \frac{2^2}{(2\pi R)^4} 
                          \Biggl[ \sum_{[n]_1}\frac{g_{8,3}^2(M_{[n]_1})}{2M_1^2}\cdot\Bigl(\frac{M_1}{M_{[n]_1}}\Bigr)^6\times4 
                          + 2\sum_{[n]_2}\frac{g_{8,3}^2(M_{[n]_2})}{2M_{[n]_2}^2}\cdot\Bigl(\frac{M_1}{M_{[n]_2}}\Bigr)^6\times 6 \times D_{[n]_2} \nonumber \\ 
                          \nonumber \\
                       && \hspace*{2.5cm} + 2^2\sum_{[n]_3}\frac{g_{8,3}^2(M_{[n]_3})}{2M_{[n]_3}^2}\cdot\Bigl(\frac{M_1}{M_{[n]_3}}\Bigr)^6\times 4 \times D_{[n]_3} 
                       + 2^3\sum_{[n]_4}\frac{g_{8,3}^2(M_{[n]_4})}{2M_{[n]_4}^2}\cdot\Bigl(\frac{M_1}{M_{[n]_4}}\Bigr)^6\times D_{[n]_4} \Biggr] (\bar{\psi}\psi)^2 
                         \nonumber \\
                       &=&\frac{1}{2} \frac{g_3^2(M_{[n]_1})}{2M_1^2} \times c_4^{(8,k=2)} \times (\bar{\psi}\psi)^2 .  
\label{4fermi-8d}
\end{eqnarray} 
We have used that the dimensionless bulk gauge coupling $\hat{g}_{8,3}$ is nearly on the UVFP, i.e.,
\begin{equation}
 \hat{g}_{8,3}^2(M_{[n]_i}) = g_{8,3}^2(M_1)\Bigl(\frac{M_1}{M_{[n]_1}}\Bigr)^4,
\end{equation}
and that
\begin{equation}
 g_{8,3}^2(M_1) = \frac{(2\pi R)^4}{2^2}g_3^2(M_1).
\end{equation}

Then the dimensionless four-fermion coupling $g_4^{(8,k=2)}$ (Eq.~(\ref{ge-4fermi})) is given by
\begin{equation}
 g_4^{(8,k=2)}=c_4^{(8,k=2)}\cdot 2^2 N_c \cdot  g_3(M_1)^2\NDA,
\end{equation}
where $c_{4}^{(8,k=2)}$ is
\begin{eqnarray}
 c_4^{(6,k=1)} &=& 4\sum_{[n]_1} \Bigl(\frac{M_1}{M_{[n]_1}}\Bigr)^6 
               + 12\sum_{[n]_2} D_{[n]_2} \Bigl(\frac{M_1}{M_{[n]_2}}\Bigr)^6 \nonumber\\
               &&+ 16\sum_{[n]_3} D_{[n]_3} \Bigl(\frac{M_1}{M_{[n]_3}}\Bigr)^6 \nonumber\\
               &&+ 8\sum_{[n]_4} D_{[n]_4} \Bigl(\frac{M_1}{M_{[n]_4}}\Bigr)^6 .
\end{eqnarray}

Next, we consider $T^4$-compactification case ($k=0$).
As in the $T^2$-case in \underline{Case.1},  the brane position does not matter and 
we take $x_5=x_6=x_7=x_8=0$: 
\begin{eqnarray}
 c_{4}^{(8,k=0)} &=& 4\times 2\sum_{[n]_1} \Bigl(\frac{M_1}{M_{[n]_1}}\Bigr)^6  
                 + 6\times 2^2\sum_{[n]_2} D_{[n]_2} \Bigl(\frac{M_1}{M_{[n]_2}}\Bigr)^6 \nonumber \\
             && + 8\times 2^3\sum_{[n]_3} D_{[n]_3} \Bigl(\frac{M_1}{M_{[n]_3}}\Bigr)^6 \nonumber \\
             && + 4\times2^4\sum_{[n]_4} D_{[n]_4} \Bigl(\frac{M_1}{M_{[n]_4}}\Bigr)^6 , 
\end{eqnarray} 
which yields
\begin{equation}
 g_4^{(8,k=0)} = c_4^{(8,k=0)}\cdot 2^2 N_c \cdot  g_3(M_1)^2\NDA. 
\end{equation}
The resultant 
$g_4$ with $c_4^{(8,k=2)},\,c_4^{(8,k=0)}$ in \underline{Case.2} is shown in  
Fig.~\ref{3brane-4fermi-1}(b).
We calculated $g_4$ till $N_{\rm KK}=100$  in this figure. 
In $T^4$-compactification case, the fermions on the 3-brane can condense which is consistent with 
Ref.~\cite{Dobrescu:1998dg, Abe:2002}, while 
for $T^4/Z_2^2$-case the $g_4$ is almost unchanged with respect to increasing $N_{\rm KK}$ 
and  
$g_4$ is always less than $g_4^{\rm crit}$ for a cutoff for the extra dimensions.
That is,
in Fig.~\ref{3brane-4fermi-1}(b) we can read $g_4^{(8)}$ as 
\begin{equation}
 g_4^{(8)}(N_{\rm KK}=100) < g_4^{\rm crit}=1 ,
\end{equation}
and hence the fermions fixed on the 3-brane do not condense by the induced four-fermion interactions
due to the bulk gluons with $T^4/Z_2^2$-compactified extra dimension.

\section{KK modes sum for  $G^{(8)}_6$}
\label{app2}
We here estimate the summation of the induced four-fermion
coupling $G^{(8)}_6$ in Eq.(\ref{eq:6dlag-GNJL7}) or its dimensionless 
coupling $g^{(8)}_6$ in Eq.(\ref{6dim4fermi8dimgluons}). 
As we discussed in the text, sum of infinite KK modes would
give us a divergent result and the anomalous dimension of the induced four-fermion
operators may make the higher KK mode contributions even more enhanced. 
But the recoil effects give us an exponential damping factor in Eq.(\ref{gauge-sup}) and 
should make the sum finite~\cite{Bando:1999}. 
Due to ignorance of the precise parameters of the exponential damping factor at this
moment, we here ignore  both the anomalous dimension effects and the recoil effects 
altogether and simply sum up finite number
of KK modes  {\it numerically} with understanding that the sum should be finite.

In the $8D \to 6D$ case,
our imposing boundary conditions are ($X=x_{\mu},x_5,x_6$)
\begin{eqnarray}
 G_M(X,x_7,x_8) &=& G_M(X,x_7+2 \pi R_7,x_8) \label{eq:bc1a}\\
 &=&G_M(X,x_7,x_8 + 2 \pi R_8), \notag \\
 G_M(X,x_7,x_8) &=& G_M(X,-x_7,-x_8). \label{eq:bc2a} 
\end{eqnarray}

The decomposition of the bulk gluons $G_M$ with $R_7=R_8=R=\Li$ is given by 
\begin{equation}
\begin{split}
 G_M (X , x_7 , x_8) = \frac{1}{2 \pi R} \Biggl[ G_{M,00}(X) &+ \sqrt{2} \sum_{n = 1}^{N_{\rm KK}} G^{[n]}_{M,c0}(X) 
\cos\frac{n x_7}{R}\\
&+ \sqrt{2} \sum_{n = 1}^{N_{\rm KK}} G^{[n]}_{M,0c}(X) \cos\frac{n x_8}{R}\\
&+ 2 \sum_{n_1,n_2 = 1}^{N_{\rm KK}} G^{[n_1,n_2]}_{M,cc}(X) \cos\frac{n_1 x_7}{\Li} \cos\frac{n_2 x_8}{R} \\
&+ 2 \sum_{n_1,n_2 = 1}^{N_{\rm KK}} G^{[n_1,n_2]}_{M,ss}(X) \sin\frac{n_1 x_7}{\Li} \sin\frac{n_2 x_8}{R} \Biggr].
\end{split} \label{eq:8d-gkk1a}
\end{equation}

In order to esitmate the upper bound for $g_6^{(8)}$, 
we calculate for $(x_{70},x_{80}) = (0,0)$ only. 
In consequence, we have
\begin{eqnarray}
 \mathcal{L}_{\rm 4F} &=& \frac{3}{4}\cdot\frac{1}{2}\cdot\frac{2}{(2\pi R)^2}\Biggl[ \sum_{[n]_1}\frac{g_{8,3}^2(M_{[n]_1})}{2M_{[n]_1}^2} \times 2 \nonumber \\
                      && \hspace*{2cm} +2\sum_{[n]_2}\frac{g_{6,3}^2(M_{[n]_2})}{2M_{[n]_2}^2} \times D_{[n]_2} \Biggr] (\bar{\psi}\psi)^2 \\
                       &=& \frac{1}{2}\frac{g_3^2(M_1)}{2 M_1^2} \times c_6^{(8)}(0,0) \times (\bar{\psi}\psi)^2 \nonumber.
\end{eqnarray}
$D_{[n]_i}$ is the number of degeneracy, that is 
the combinations of $(n_1,n_3, \cdots)$ having the same KK-gluons masses:
$M_{[n]_i}(M^2_{[n]_i}= (n_1^2 + n_2^2 + \cdots) M_1^2)$ 
where $M_1=\Lambda$.

Next, we have used the fact that dimensionless bulk gauge coupling $\hat{g}^2_{8,3}$ is approximately near the UVFP
and set 
\begin{equation}
 g_{8,3}^2(n\Lambda)=\frac{g_{8,3}^2(\Lambda)}{n^4}.
\end{equation}
Thus considering $g^2_{6,3}(\Lambda) \equiv 2 g^2_{8,3}(\Lambda)/{(2 \pi R)^2}$, 
we get the total coefficient of four-fermion operator 
\begin{equation}
 c_6^{(8)}(0,0) = 2\sum_{[n]_1} \Bigl(\frac{M_1}{M_{[n]_1}}\Bigr)^6 
               + 2\sum_{[n]_2} D_{[n]_2} \Bigl(\frac{M_1}{M_{[n]_2}}\Bigr)^6,
\end{equation}

Hence the bound of the dimensionless induced four-fermion coupling defined in Eq.(\ref{dless-4coupling0}) is given by
\begin{eqnarray}
 g_6^{(8)} &=& c_6^{(8)}(x_{70},x_{80})\cdot 2^2 N_c \cdot \hat{g}_{6*,3}^2\NDA = 
 c_6^{(8)}(x_{70},x_{80})\cdot 2^2 N_c \cdot \frac{3}{44} \nonumber \\
 &=& c_6^{(8)}(x_{70},x_{80})\cdot\frac{3 N_c}{11} \\
 &\le & c_6^{(8)}(0,0)\cdot\frac{3 N_c}{11}
\end{eqnarray}
where we again used $\hat{g}_{6,3}^2\NDA=\hat{g}_{6*,3}^2\NDA=3/44$.

\begin{figure}
\begin{center}
 \includegraphics[width=0.5\hsize,clip]{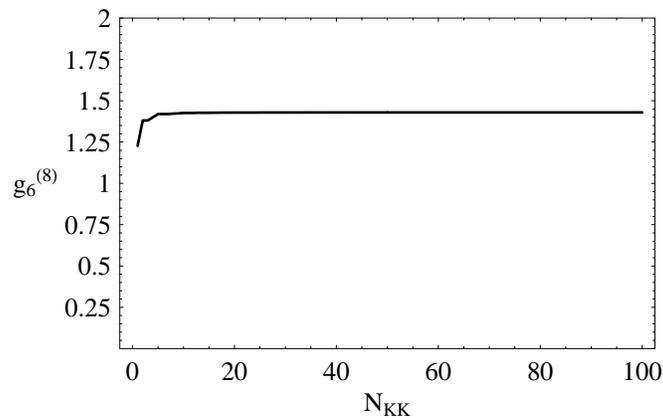}
\end{center}
\caption{The estimation of $g^{(8)}_6$ for Eq.~(\ref{eq:6dlag-GNJL7})
         for the summation by $N_{\rm KK}$(KK-mode $\text{mass}^2$ is $N_{\rm KK}^2M_1^2$)
         }
\label{8-6-sum1}
\end{figure}

The numerical calculation result for the upper bound of $g_6^{(8)}$
(sum by $N_{\rm KK} = 100$) is shown in Fig.~\ref{8-6-sum1},
that is, we get the upper bound of $g_6^{(8)}$:
\begin{equation}
 g_6^{(8)} \lesssim 1.42.
\end{equation}
Since this upper bound is nearly the same as the one in  Eq.~(\ref{4th}), the sum till the
4th KK modes,
we may conclude that all KK-modes effects contributions are well approximated by
the first few KK-modes effects contributions.
If we consider recoil effects more seriously, the main contribution may even be the lowest KK-mode only.

\end{document}